\documentclass[intlimits,twoside,a4paper]{article}

\usepackage[cp1251]{inputenc}

\usepackage[eqsecnum]{cmpj3}

\usepackage{bm}

\issue{2024}{27}{2}{23201}
\doinumber{10.5488/CMP.27.23201}

\title[Water-ethanol liquid mixtures]
{Molecular dynamics simulations of water-ethanol mixtures.
I. Composition trends in thermodynamic properties
}

\author[D. Benavides Bautista, M. Aguilar, O. Pizio]
{D. Benavides Bautista\refaddr{label1},
M. Aguilar\orcid{0000-0003-3850-1188}\refaddr{label2}, 
        O. Pizio\orcid{0000-0001-8333-4652}\refaddr{label2}
\thanks{Corresponding author: \email{oapizio@gmail.com}.}}

\addresses{
\addr{label1} Instituto de Ciencias B\'asicas y Ingenier\'ia, 
Universidad Aut\'{o}noma de Estado de Hidalgo, Pachuca de Soto, Hidalgo 42039,
M\'{e}xico
\addr{label2}Instituto de de Qu\'{i}mica, Universidad Nacional Aut\'{o}noma de M\'{e}xico,
Circuito Exterior, 04510, Cd. Mx., M\'{e}xico}

\Keywords{molecular dynamics, water-ethanol mixtures, surface tension, dielectric constant, partial
molar volumes}

\date{Received April 16, 2024, in final form May 20, 2024}

\begin{document}

\maketitle

\begin{abstract}
We explored the composition dependence of a rather comprehensive set of
properties of liquid water-ethanol mixtures
by using the isobaric-isothermal molecular dynamics computer simulations.
The united atom non-polarizable model from the TraPPE
data basis for the ethanol molecule 
combined with the TIP4P-2005 and SPC/E  water 
models is considered. 
We restrict our calculations to atmospheric pressure, 0.1013 MPa, and room temperature, 298.15~K.
Composition trends of the behavior of density, excess mixing volume,
apparent molar volumes are described. On the other hand, the excess mixing 
enthalpy and partial molar enthalpies of species are reported. 
Besides, we explore  the coefficient of isobaric thermal expansion,
isothermal heat capacity, adiabatic bulk modulus and heat capacity 
at constant pressure. 
In addition, the self-diffusion coefficients of species, 
the static dielectric constant and the surface tension are described.
We intend to get insights into  peculiarities of mixing of species in the mixture 
upon changes of ethanol molar fraction. The quality of predictions of the models 
is critically evaluated by detailed comparisons with experimental results.
Then, necessary improvements of the modelling are discussed.

\printkeywords
%
%
\end{abstract}

\section{Introduction}

One of us (O.P.) with profound sadness would like to dedicate this work to the memory of 
his former scientific supervisor and
founder of the Condensed Matter Physics journal, Prof. Ihor Yukhnovskyi, who
passed away  recently.

This manuscript presents the results of the first part of our project focused on the
properties of water-ethanol mixtures. It refers to the description of the composition dependence
of thermodynamic and other related issues of mixing of water and ethanol species
at room temperature and at ambient pressure. To do that we use molecular dynamics 
computer simulations. All aspects of the evolution of the microscopic structure
and of hydrogen bonds network will be discussed in future, in our subsequent report.

Liquid mixtures of water and ethanol are of much practical importance as solvents
and reaction media in organic chemistry, medicinal and food chemistry, and in
chemical engineering.
Pure ethanol and its mixtures with water have 
been investigated by several experimental techniques since the times of
Mendeleev.
The experimental knowledge and understanding
of the microscopic structure and dynamic properties of the systems in question
mainly follow from the application of  neutron scattering,
nuclear magnetic resonance, dielectric relaxation, vibrational 
and Raman  spectroscopy methods~\cite{stewart,raman,harris,petong}.
On the other hand, calorimetric studies yield a valuable set of thermodynamic data 
for the mixtures in question~\cite{benson,costigan,koga1,koga2,grolier}.
Dynamic light scattering studies
contributed to the elucidation of anomalous behaviors in
water-ethanol mixtures as well, see, e.g.,~\cite{kojima} and 
references in~\cite{chechko1,chechko2}.

In order to interpret the experimental observations in every detail 
and to get ampler insights, one is usually
forced to resort to computer simulation methodology.
A common strategy of computer simulations methodology is to choose a model of each species, 
water and ethanol in the present case,
and assume the cross interactions by using the combination rules. 
Then, software based on strict rules of statistical mechanics is applied.
Appropriateness of the computer simulations predictions for a given model for
a mixture, upon changing temperature, $T$, pressure, $P$, and composition, $X$,
variables, is then tested by comparison with reference experimental data.  

Profound insights into the properties of pure components of interest
of the present study from computer simulations are available.
Namely, a comprehensive set of data for non-polarizable 
water models were provided by Vega and Abascal~\cite{vega-pccp}.
A similar type of strategy of description was applied to
methanol~\cite{salgado}. In the case of ethanol and higher alcohols,
the situation is less satisfactory.
We are not aware of the work describing an ample set of ethanol properties
using different models with their critical evaluation versus experimental data.
Certain aspects of the microscopic structure and
some properties of pure ethanol were studied in \cite{guardia1,guardia2,zangi}.
On the other hand, many computer simulation studies were 
performed for water-ethanol mixtures. They differ in the modelling of water component.
Specifically, the TIP3P water model was used in~\cite{camp}. 
Wensink et al.\cite{wensink} used the TIP4P water model.
However, most frequently well tested SPC/E~\cite{spce} and TIP4P-2005~\cite{carlos} models 
were used to describe
water species~\cite{sokolic2011,sokolic2015,sokolic2016,pusztai2015,pusztai2016,pusztai2018}.
Concerning ethanol species, various models were involved. 
All of them have roots in the OPLS developments of Jorgensen et al.~\cite{jorgensen1,jorgensen2}.
Namely, the works from the Croatian laboratory explored versions of the
united atom model for ethanol~\cite{sokolic2015} (the CH$_3$ and CH$_2$ groups are considered
as sites) including the TraPPE version~\cite{trappe}. 
The works from the Hungarian laboratory of L. Pusztai focused on
the OPLS all atom models~\cite{pusztai2015}. Similar modelling was used
by Wensink et al.~\cite{wensink}. The united atom type models were used in~\cite{vrabec,bagchi}.
Finally, the attempt to study water-ethanol mixture
using  polarizable models for two components was undertaken as well~\cite{patel}.
The united site models are computationally less time-consuming, in comparison with more
sophisticated ones. They can be used, however, with a certain degree of confidence,
to elucidate the mixing trends of species and to describe thermodynamic properties.
On the other hand, these models have some disadvantages; since not all the atoms 
are represented, they miss some of the partial radial distribution functions and 
stereochemical information, such as some bond angles and dihedral angles. 
Therefore, if one focuses on the microscopic structure in terms of the total structure factor
to compare with experimental results, it is appropriate to perform simulations
with all-atom models, see, e.g.,~\cite{galicia,pusztai2015}.

Having this discussion in mind, the principal objective of the present work is 
to obtain a rather comprehensive set of results 
for the composition dependence of water-ethanol mixtures at 
atmospheric pressure and at room temperature by using isobaric-isothermal molecular 
dynamics computer simulation. 
The reason is that several previous publications were focused on
the interval of low ethanol concentration. Namely, elucidation of
anomalies of composition behavior, or better say of non-monotonous behaviors of different 
properties and their interpretation from molecular dynamics simulations,
is explored. Moreover, a restricted set of properties
of the systems in question is usually considered. Consequently, the validation of the models 
used in computer simulations is not complete.


\section{Models and simulation details}

In this work we restrict our attention to a single united atom type, non-polarizable model 
with four sites, O, H, CH$_2$, CH$_3$ for ethanol~\cite{trappe}.
Within this type of modelling, the interaction  potential between all atoms and/or groups is assumed 
as a sum of Lennard-Jones (LJ) and Coulomb terms. The parameters are given in the 
web page of the TraPPE data basis (\url{http://trappe.oit.umn.edu)} and in the original 
publication~\cite{trappe}.
For water, the TIP4P-2005~\cite{carlos} and SPC/E~\cite{spce} models are used. 
Lorentz-Berthelot combination rules were used to determine the cross parameters for
the relevant potential well depths and diameters.

Molecular dynamics computer simulations of water-ethanol mixtures are performed in the
isothermal - isobaric (NPT) ensemble at atmospheric  pressure 1~bar and at temperature 298.15 K.
We used GROMACS package~\cite{gromacs} version 5.1.2.
The simulation box in each run was cubic, the total number of molecules of both species is
fixed at 3000. Composition of the mixture is described by the molar fraction of ethanol
molecules, $X_{\text{eth}}=N_{\text{eth}}/(N_{\text{eth}}+N_w)$.
As common, periodic boundary conditions were used.
Temperature and pressure control was provided by the V-rescale thermostat and Parrinello-Rahman
barostat with $\tau_T$ = 0.5 ps and $\tau_P$ = 2.0 ps, the timestep was 0.002 ps.
The value of $4.5\cdot 10^{-5}$~bar$^{-1}$ was used for the compressibility of mixtures.

The non-bonded interactions were cut-off at 1.1 nm, whereas the long-range electrostatic interactions
were handled by the particle mesh Ewald method implemented in the GROMACS software package  (fourth
order, Fourier spacing equal to 0.12) with the precision $10^{-5}$.
The van der Waals correction terms to the energy and pressure were used.
In order to maintain the geometry of water molecules and ethanol intra-molecular bonds rigid, the LINCS
algorithm was used.

After preprocessing and equilibration, consecutive simulation runs, each for not less than 10 ns,  with
the starting configuration being the last configuration from the previous
run, were performed to obtain trajectories for the data analysis. 
The results for each property  were obtained by averaging over 7--10 production runs. 

\section{Results and discussion}

\subsection{Density of ethanol-water mixtures on composition}

As we mentioned in the introductory section, there were several
experimental reports concerning the density of water-ethanol  mixtures upon changing composition. 
We used experimental data at room temperature $T= 298.15$~K, and at 
atmospheric pressure~\cite{pecar,hervello}.

\begin{figure}[h]
\begin{center}
\includegraphics[width=6.5cm,clip]{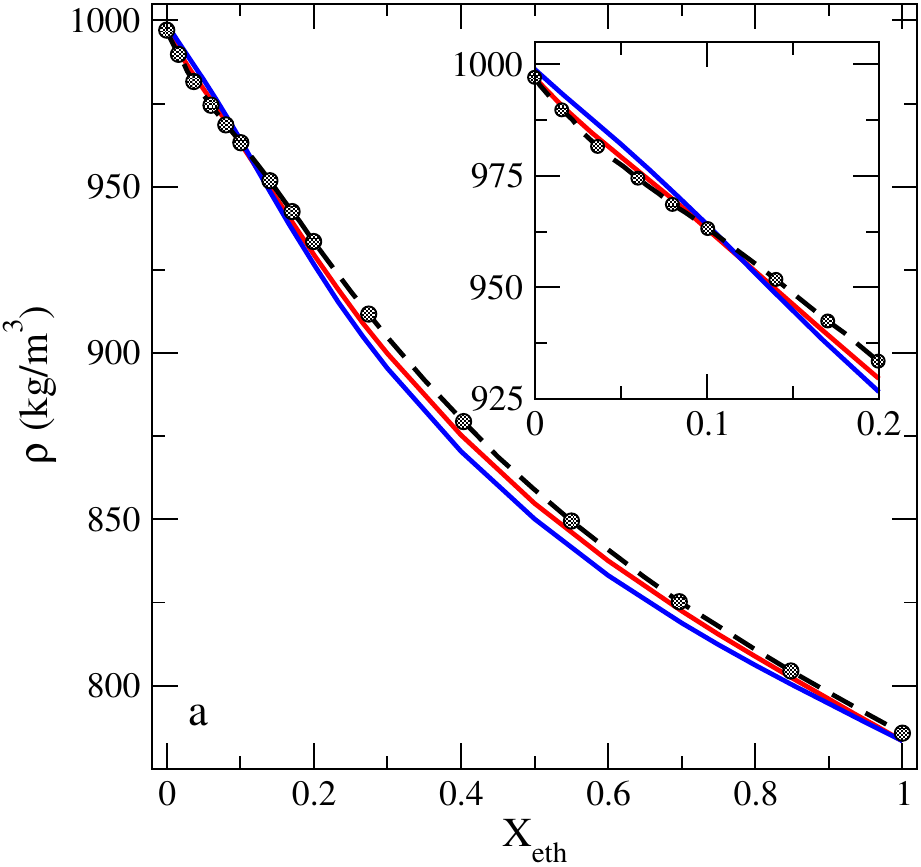}
\includegraphics[width=6.5cm,clip]{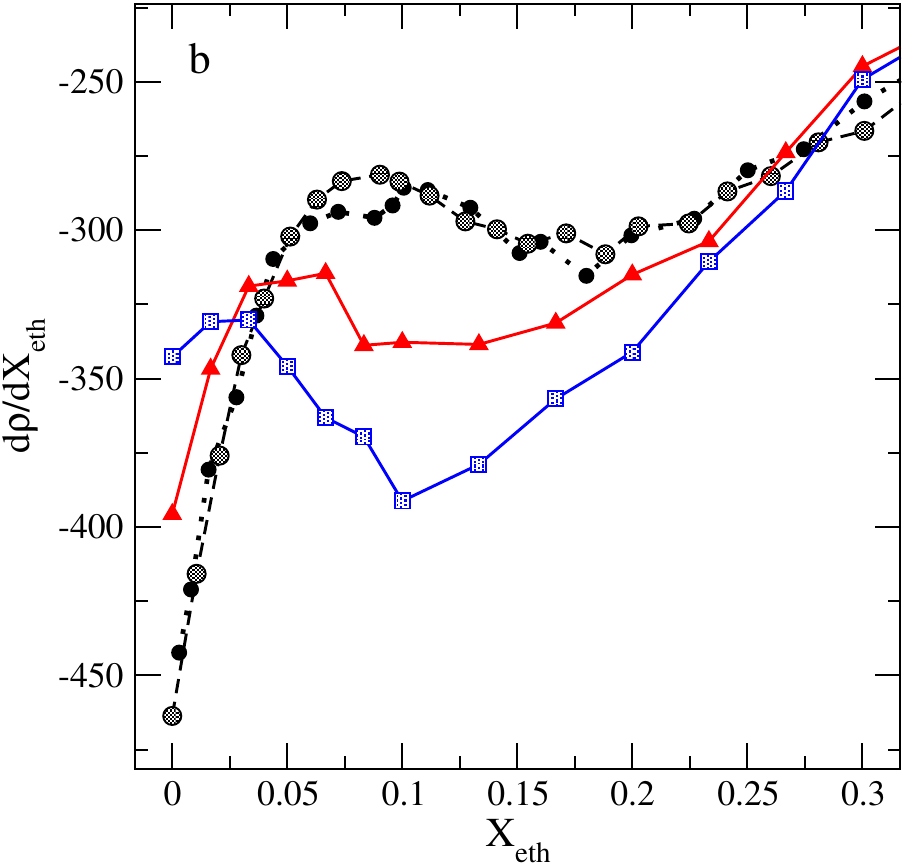}
\end{center}
\caption{(Color online) Panel a: Composition dependence of density of water-ethanol
mixtures from the NPT MD simulations of TIP4P/2005-TraPPE model (solid red line),
SPC/E-TraPPE model (solid blue line),
in comparison  with the experimental data (dashed lines with circles) 
at $T = 298.15$~K~\cite{pecar}. The inset provides an enhanced view of density changes at low
ethanol concentrations.
Panel b: Derivative of density by ethanol concentration at low $X_{\text{eth}}$, experimental
data from \cite{pecar} --- circles with dotted line, \newline\cite{hervello} --- circles with dashed
line, TIP4P/2005-TraPPE model --- red triangles, SPC/E-TraPPE --- blue squares.
}
\label{fig1}
\end{figure}

Both combinations of models of this study describe the dependence of density on $X_{\text{eth}}$
quite well. Pure ethanol density is $\approx 783.5$~kg/m$^3$ close to the experimental 
value 785.2  \cite{pecar} or 785.7 kg/m \cite{hervello}. The most pronounced deviation
of simulation data from experimental results are observed at intermediate compositions,
but inaccuracy is not big, however (figure~\ref{fig1}a). At a low ethanol concentration, both models capture the
maximum of the derivative of density on $X_{\text{eth}}$, figure~\ref{fig1}b. This behavior 
describes the peculiarity of mixing of a small amount of ethanol in the medium
of water species. It witnesses the contraction of the mixture volume and is commonly attributed to 
the hydrophobic effect. We discuss this issue  more in detail herein below.

\subsection{Excess mixing volume} \label{sec3.2}

It is important not only to describe the trends of behavior of a given property
on composition, but to capture correctly the deviation from ideality as well.
These insights follow, for example,  from the behavior of the excess
density or the excess mixing volume. The excess mixing volume is defined as 
follows, $\Delta V_{\text{mix}} = V_{\text{mix}} - X_{\text{eth}} V_{\text{eth}} - (1-X_{\text{eth}}) V_w$,
where $V_{\text{mix}}$, $V_{\text{eth}}$ and $V_w$ refer to the molar volume of the mixture and
of the individual components, ethanol and water, respectively.

\begin{figure}[h]
\begin{center}
\includegraphics[width=7cm,clip]{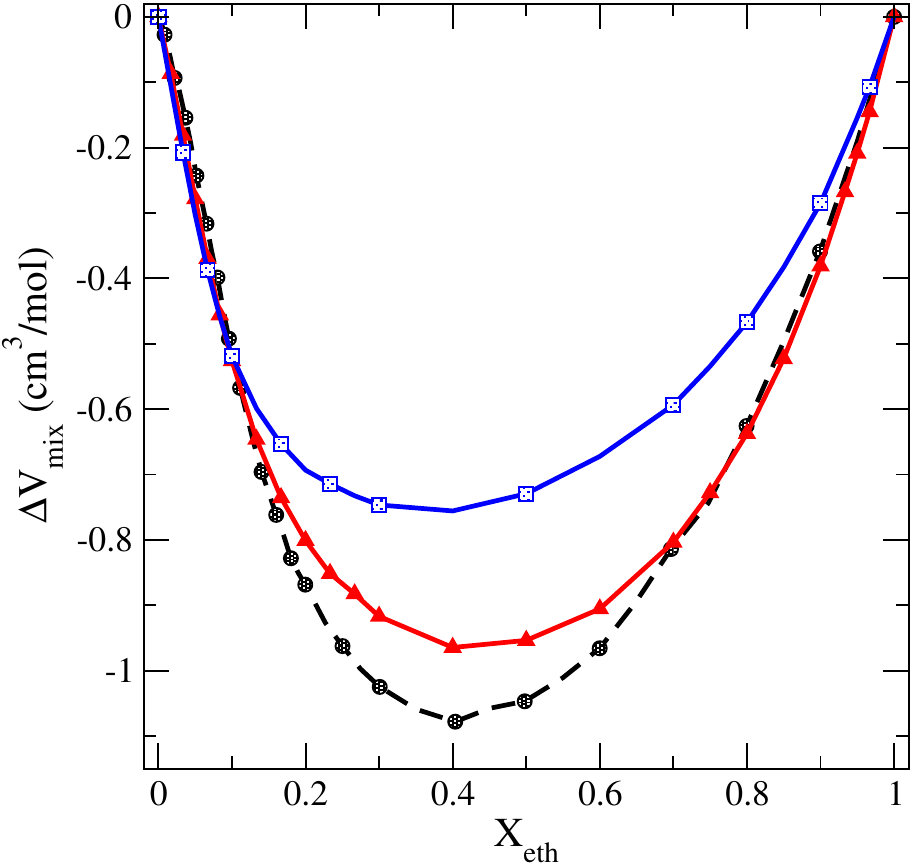}
\end{center}
\caption{(Colour online) A comparison of the composition dependence of the excess mixing volume
of water-ethanol mixtures for models as in figure~\ref{fig1} with
the experimental data of \cite{pecar}. The experimental data are
given by black dashed line with circles; red line with triangles and
blue line with squares correspond to
TIP4P/2005-TraPPE model and SPC/E-TraPPE model, respectively.
The results refer to  298.15~K and atmospheric pressure.
\label{fig2}
}
\end{figure}

Experimental data show that $\Delta V_{\text{mix}}$ is negative and exhibits
a minimum at $X_{\text{eth}} \approx 0.4$, figure~\ref{fig2}.
The simulation results show qualitatively similar trends of behavior.
A comparison between the experiment and simulations 
with TIP4P-2005-TraPPE model can be termed as quite satisfactory.
The SPC/E-TraPPE model is less accurate
concerning the description of  geometric aspects of mixing of species.
In this projection of the equation of state,
a single peculiar point on composition is well observed.

\begin{figure}[h]
\begin{center}
\includegraphics[width=6.0cm,clip]{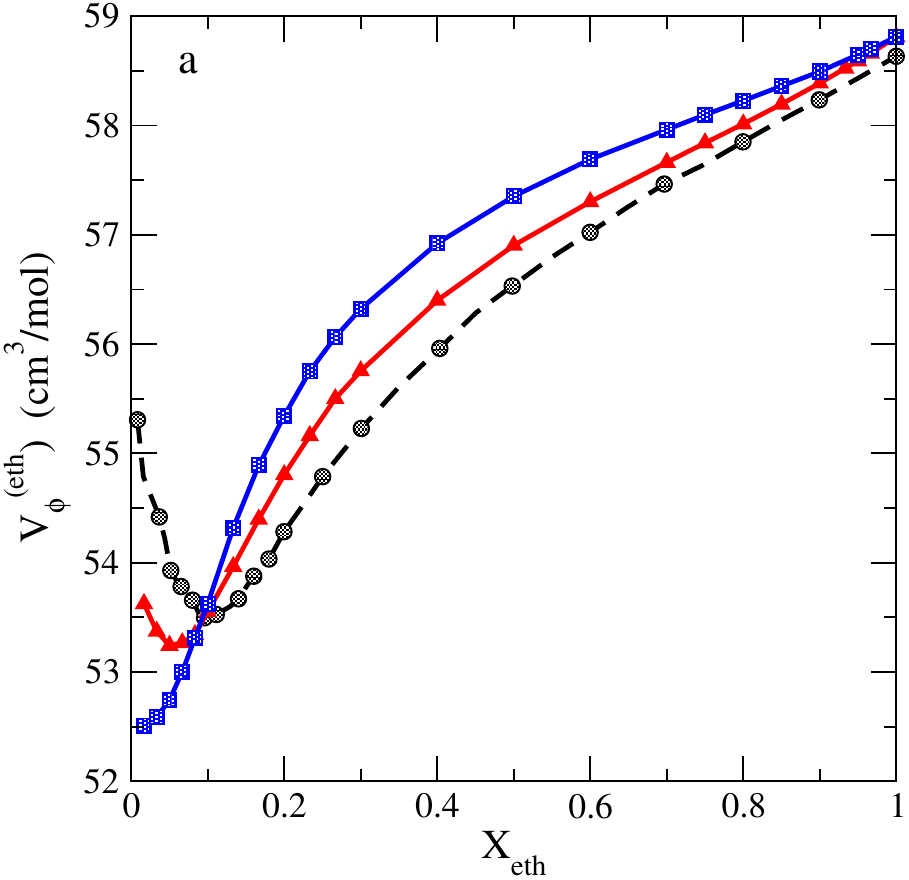}
\includegraphics[width=6.0cm,clip]{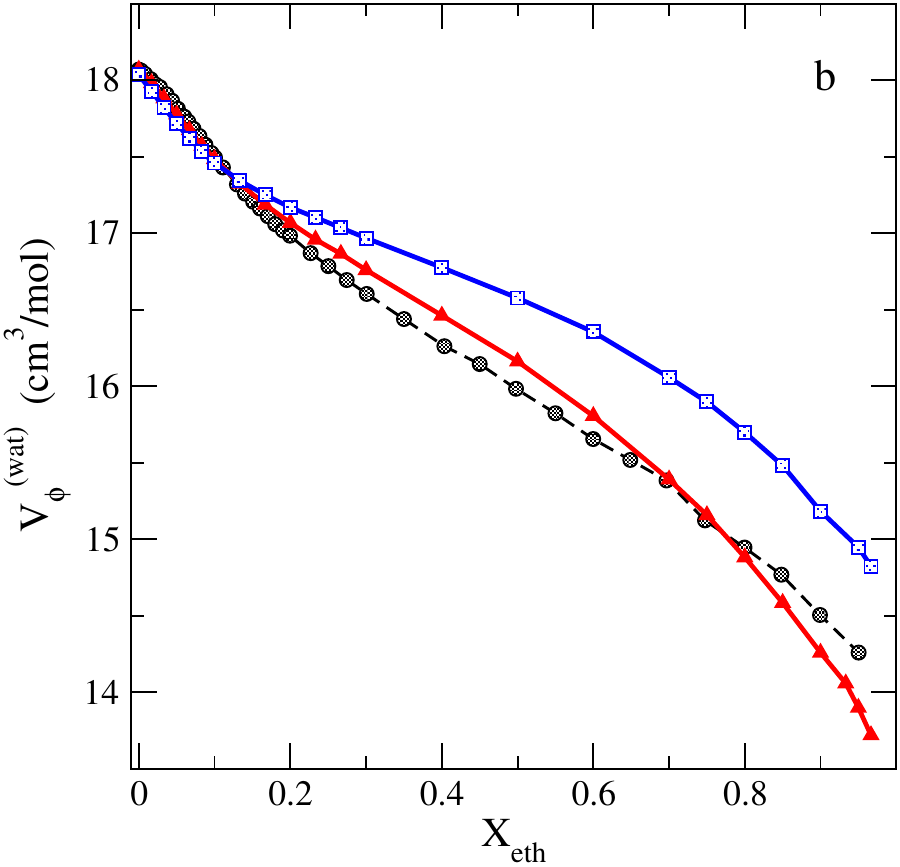}
\end{center}
\caption{(Colour online) 
A comparison of the composition dependence of the
apparent molar volumes of ethanol and water species from simulations,
with the experimental data~\cite{pecar} at 298.15~K. The nomenclature of
lines and symbols as in figure~\ref{fig2}.
\label{fig3}
}
\end{figure}

In order to discern the contributions of each species into the excess molar volume 
and to obtain deeper insights into the geometric aspects of mixing on composition,
both from experiments and simulations, one can resort to the notion of
the apparent molar volume of species rather than the excess molar volumes.
The apparent molar volume
for each species according to the definition is \cite{torres}:
$V_{\phi}^{(w)}= V_w + \Delta V_{\text{mix}}/(1-X_{\text{eth}})$ 
and  $V_{\phi}^{(\text{eth})}= V_{\text{eth}} + \Delta V_{\text{mix}}/X_{\text{eth}}$.
We elaborated the experimental density data from \cite{pecar}
and the results from our simulations to construct the plots shown in panels a and b of figure~\ref{fig3}.
These plots confirm that the TIP4P-2005-TraPPE model provides a quite reasonable description of the
composition behavior for $V_{\phi}^{(\text{eth})}$ in water-rich mixtures in the entire composition range.
The minimum of  $V_{\phi}^{(\text{eth})}$ is predicted at a slightly
lower ethanol concentration, $X_{\text{eth}} \approx 0.0667$, in comparison with the experimental
result, $X_{\text{eth}} \approx 0.1$.   
On the other hand, the SPC/E-TraPPE model is less accurate in this respect. Apparently,
the minimum of $V_{\phi}^{(\text{eth})}$ exists at a much lower ethanol concentration
for this model. Concerning the apparent molar volume of water (panel b of figure~\ref{fig3}), 
we observe that the TIP4P-2005-TraPPE model leads to more accurate predictions.

This kind of composition behavior of apparent molar volumes of ethanol species
in water-rich mixtures can be related to the experimental results for abnormal
intensity of scattered light~\cite{kojima,chechko1,chechko2}. 
Experimental evidence of a peculiar point at $X_{\text{eth}} \approx 0.12$ corresponding
to the largest concentration fluctuations is commonly interpreted in terms of
the formation of ethanol clusters and changes of their shape.  
From the thermodynamic point of view, the minimum of the apparent
molar volume of ethanol indicates the hydrophobic effect at this composition interval.

It is worth mentioning that some combinations of alcohol and water models,
in spite of apparently accurate description of density and 
the excess mixing volume, do not capture the minimum of the excess
apparent volume of alcohol and do not correctly reproduce the temperature dependence of
hydrophobic effect~\cite{mario}. This issue requires additional
studies for water-ethanol mixtures. 

\begin{figure}[h]
\begin{center}
\includegraphics[width=6.0cm,clip]{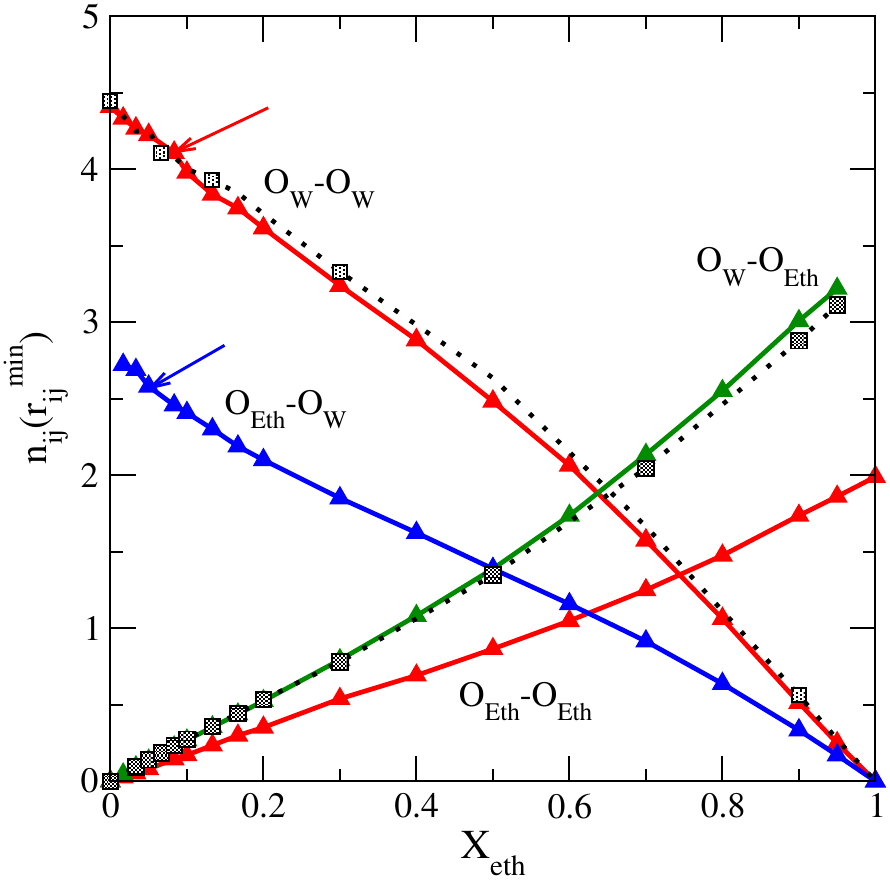}
\includegraphics[width=6cm,clip]{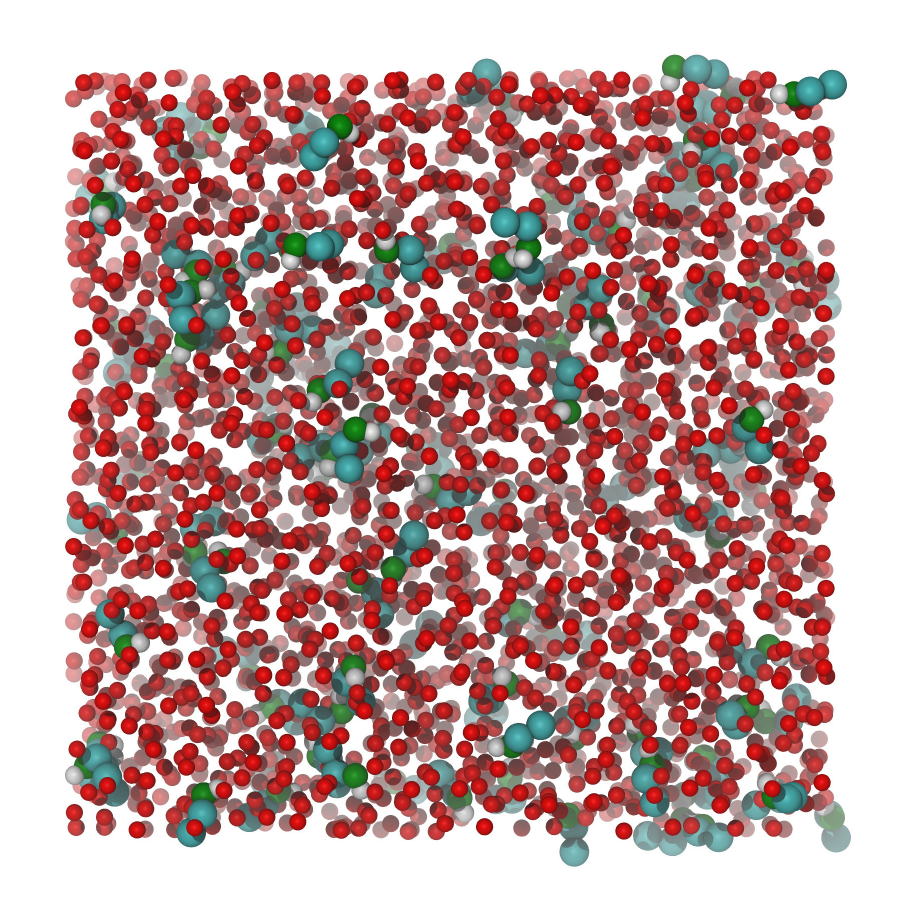}
\end{center}
\caption{(Colour online) Left-hand panel: composition dependence of the first coordination number of
oxygens belonging to water and to ethanol species at 298.15~K.
The solid lines with symbols and dotted line with symbols correspond to
TIP4P/2005-TraPPE and SPC/E-TraPPE model, respectively.
Right-hand panel: a snapshot of ethanol molecules typical configuration in
water medium (red small circles - O$_w$) at $X_{\text{eth}} = 0.0333$.
\label{fig4}}
\end{figure}

The minimum of the ethanol excess apparent molar volume is observed at small 
value for $X_{\text{eth}}$, at $X_{\text{eth}} \approx 0.0667$ for TIP4P/2005-TraPPE model.
At such conditions, the water-ethanol mixture has got the density quite close to
the one of pure water. In order to get insight into the 
mutual distribution of particles in such mixtures, we performed calculations
of the first coordination number of ethanol and water oxygens.
The first coordination number is commonly evaluated  by integration
of the radial distribution functions up to the first minimum, 
as follows,
\begin{equation} 
  n_{ij} = 4\piup \rho_j \int_0 ^{r^{\text{min}}_{ij}}  g_{ij}(R)  R^2 \rd R,
\end{equation}
where $\rho_j$ is the density of species $j$ and  $g_{ij}(r)$ is the corresponding 
pair distribution function. The anomaly of $V_{\phi}^{(\text{eth})}$ occurs
in highly coordinated mixtures. Namely, the coordination number of water oxygens
is $\approx$4 whereas the ethanol oxygen is surrounded by $\approx$ 2.5 water
oxygens. Moreover, these two first coordination numbers exhibit weak peculiarity
at low $X_{\text{eth}}$ values (marked by arrows in figure~\ref{fig4}, left-hand panel). 
The snapshot of the configuration of ethanol molecules at $X_{\text{eth}} \approx 0.033$
in right-hand panel of figure~\ref{fig4}, illustrates that their distribution is not entirely uniform. 
Microheterogeneity of the distribution of particles upon changing the composition
of the mixture is the subject of many previous studies~\cite{sokolic2015,
sokolic2016,bagchi}. This issue is out of scope of the present report, however.

\subsection{Energetic aspects of mixing of ethanol and water molecules}

\begin{figure}[h]
\begin{center}
\includegraphics[width=6.5cm,clip]{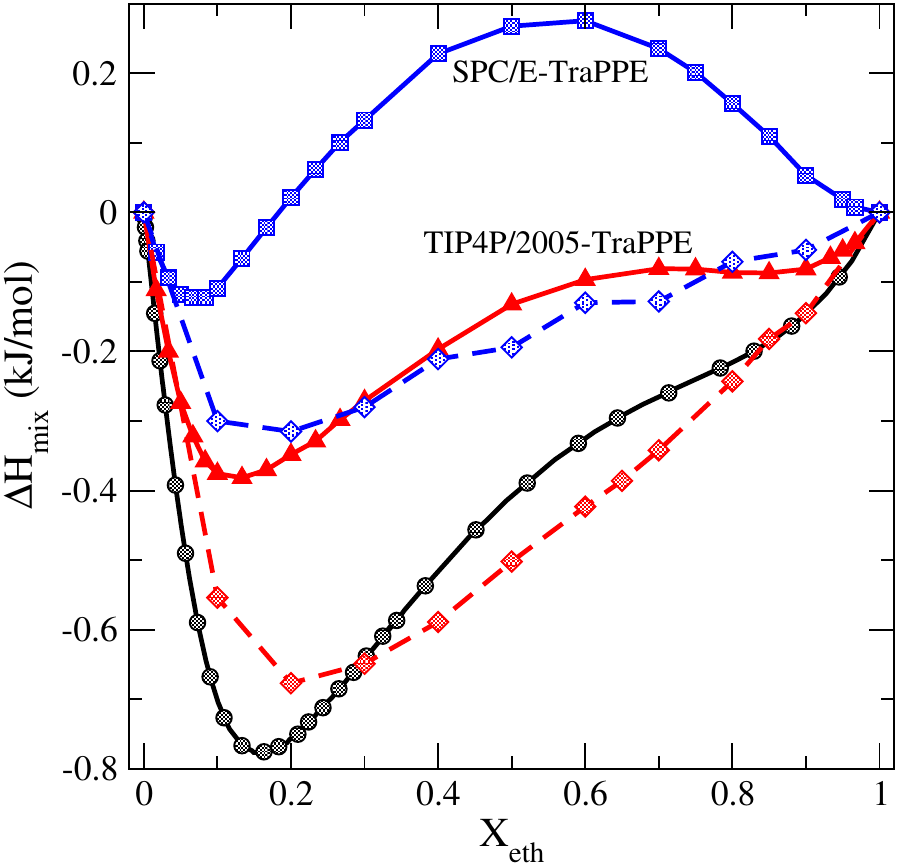}
\end{center}
\caption{(Colour online) A comparison of the behavior of the excess mixing enthalpy
from simulations and experimental data from \cite{costigan}.
The nomenclature of lines and symbols is the same as in figure~\ref{fig2}.
Two additional curves reproduce the results from \cite{vrabec}
with TIP4P/2005 water model (red diamonds) and with SPC/E model (blue diamonds)
combined with ethanol model of their own design.
\label{fig5}
}
\end{figure}

Energetic manifestation of mixing trends is commonly discussed in terms of
excess mixing enthalpy. We used the experimental results from
\cite{costigan,koga1,koga2} and our simulation data to explore the mixing enthalpy 
upon composition of water-ethanol mixtures. The results are given in figure~\ref{fig5}.

There, we observe that the TIP4P-2005-TraPPE model qualitatively reproduces the shape of
experimental behavior. The minimum location and peculiarity of $\Delta H_{\text{mix}}$  
at high values of $X_{\text{eth}}$
agree reasonably well  with the experimental trends.
The absolute values from  TIP4P-2005-TraPPE model
for $\Delta H_{\text{mix}}$ are underestimated, however. The 
SPC/E-TraPPE model predictions are not satisfactory for $\Delta H_{\text{mix}}$.
Large scattering of data for $\Delta H_{\text{mix}}$ upon composition for water-ethanol mixtures
is comprehensively documented in figure~2 of~\cite{sokolic2015} and
in figure~22 of~\cite{vrabec}. Moreover, the absolute values for $\Delta H_{\text{mix}}$
from the study of Wensink et al.~\cite{wensink} are almost twice larger than the
experimental ones.
In order to elucidate the reasons of this kind of discrepancy of modelling and
experiment, it is worth to resort to the partial excess molar enthalpies.
They result from the excess mixing enthalpy, $\Delta H_{\text{mix}}$ as
follows~\cite{torres},
\begin{equation}
 h_w^{\text{ex}} =\Delta H_{\text{mix}}+X_{\text{eth}}\left.\left(\frac{\partial \Delta H_{\text{mix}}}{\partial X_w}\right)\right|_{P,T}\,,
\end{equation}
\begin{equation}
 h_{\text{eth}}^{\text{ex}} =\Delta H_{\text{mix}} -X_w\left.\left(\frac{\partial \Delta H_{\text{mix}}}{\partial X_w}\right)\right|_{P,T},
\end{equation}
where, $X_w = 1- X_{\text{eth}}$.

\begin{figure}[h]
\begin{center}
\includegraphics[width=6.5cm,clip]{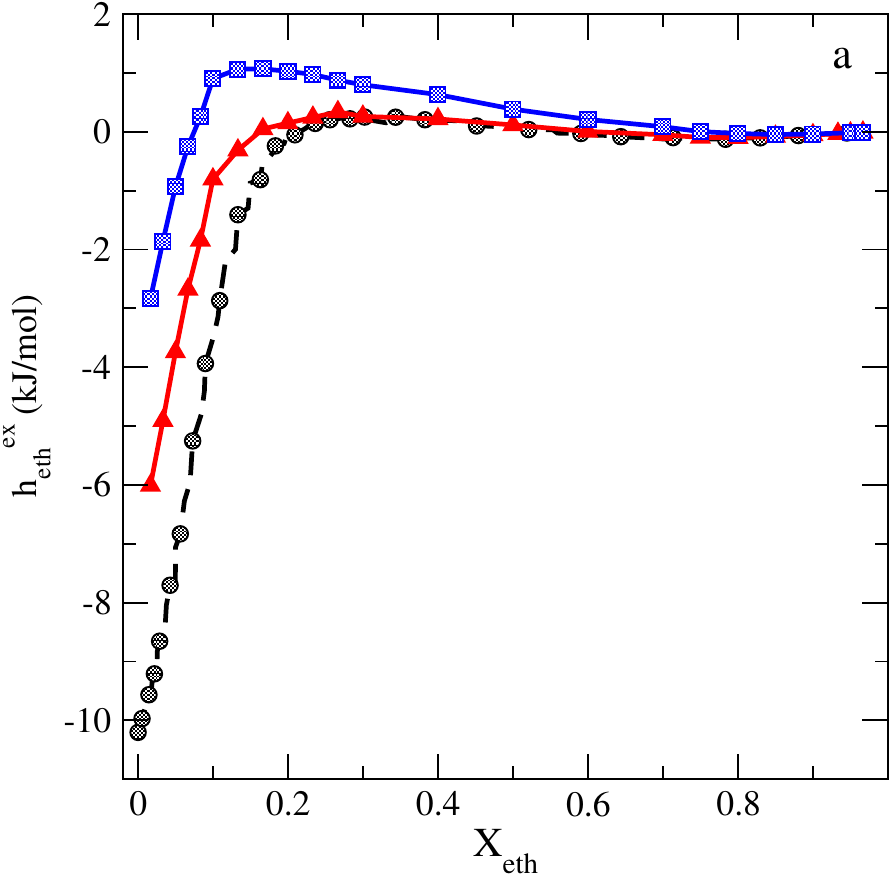}
\includegraphics[width=6.5cm,clip]{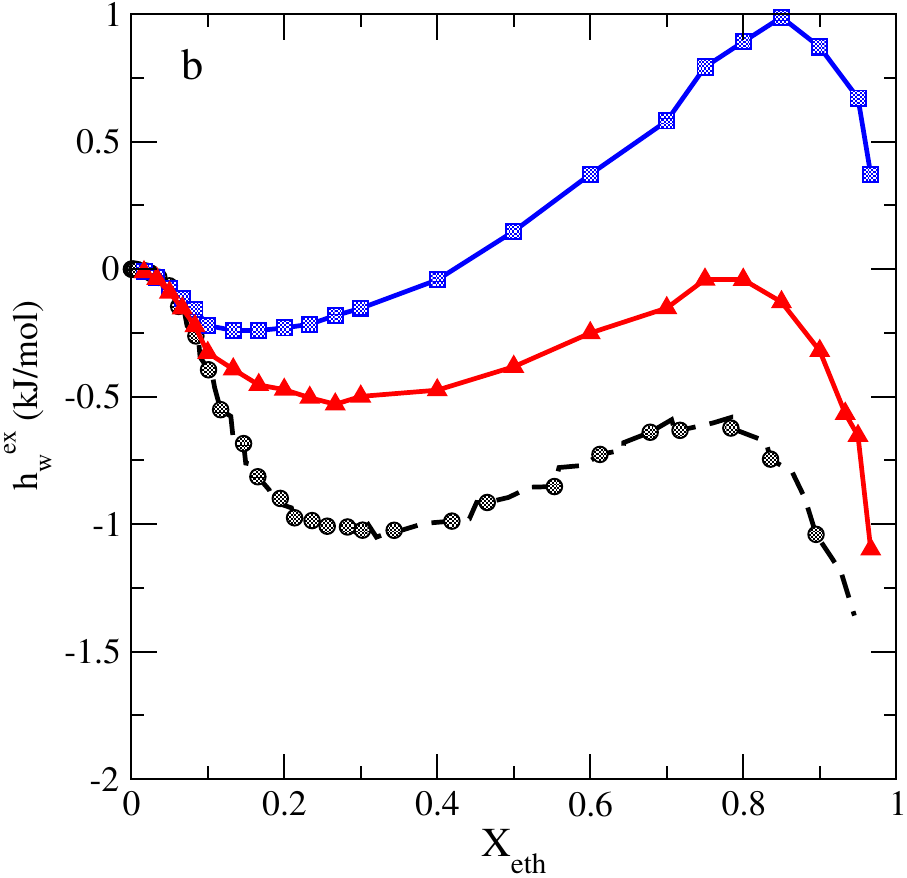}
\end{center}
\caption{(Colour online) A comparison of the behavior of the partial molar excess enthalpy
from simulations and experimental data from \cite{koga1,koga2}.
The nomenclature of lines and symbols is the same as in figure~\ref{fig2}.
\label{fig6}
}
\end{figure}

In general terms, the computer simulation predictions for $h_{\text{eth}}^{\text{ex}}$ are reasonable,
as we see from figure~\ref{fig6}a. On the other hand, the shape of behavior of $h_w^{\text{ex}}$ reproduces the
experimental behavior as well, figure~\ref{fig6}b. 
However, the absolute values of this property substantially differ from the 
experimental data. Still, the TIP4P-2005-TraPPE model provides a satisfactory 
description of the energetic trends of mixing of ethanol and water species. 

\subsection{On the predictions from fluctuations}

Now, we would like to turn our attention to a set of properties that
proceed from fluctuations. These quantities and their composition dependence 
are less frequently discussed in literature compared to the excess mixing volume 
and enthalpy and require more computational efforts, see, e.g.,~\cite{jose} for pure water.  

\begin{figure}[h]
\begin{center}
\includegraphics[width=6.5cm,clip]{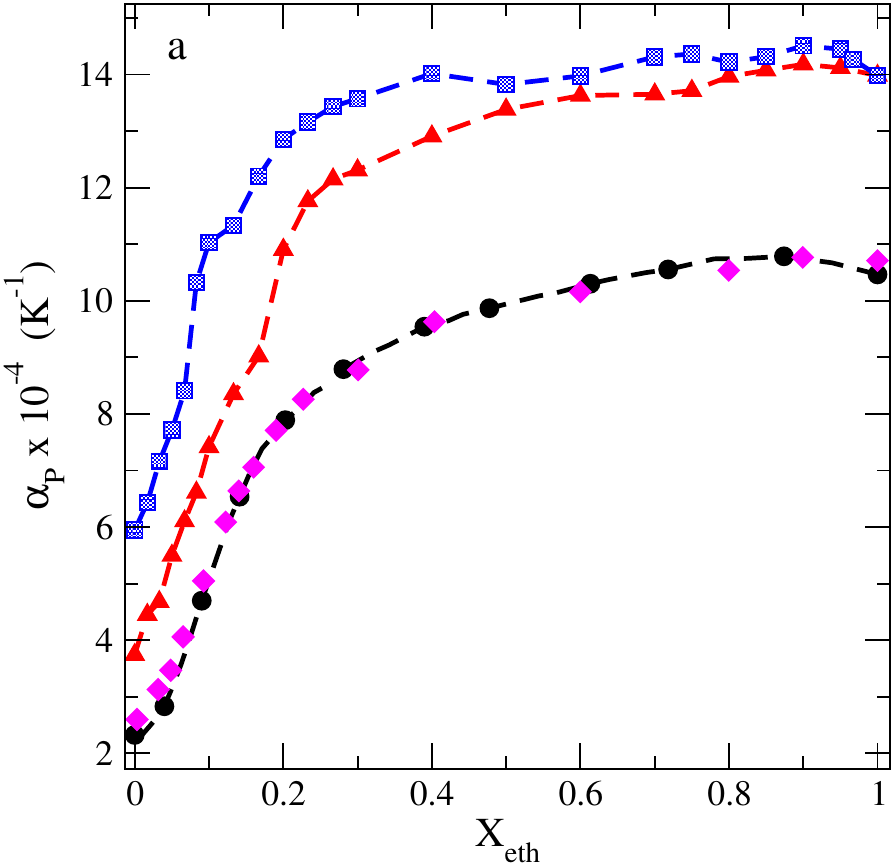}
\includegraphics[width=6.5cm,clip]{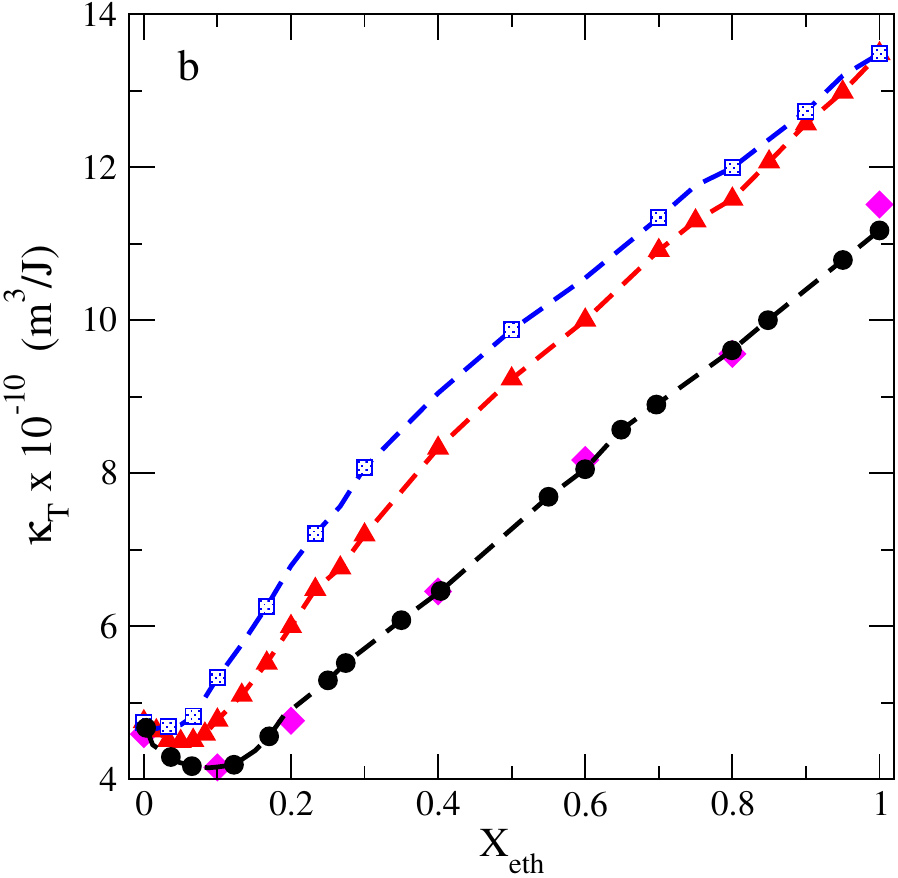}
\end{center}
\caption{(Colour online) A comparison of the composition dependence of the 
coefficient of the isobaric thermal expansion, $\alpha_P$, and
isothermal compressibility, $\kappa_T$, in panels a and b, respectively,
with the experimental data. Panel a: circles and diamonds 
are from~\cite{pecar} and from~\cite{hervello}, respectively.
Panel b: circles and diamonds
are from~\cite{pecar} and from~\cite{tanaka}, respectively.
\label{fig7}
}
\end{figure}

The thermal expansion coefficient, $\alpha_P$,  and isothermal compressibility,  $\kappa_T$,

\begin{equation}
\alpha_P = \frac{1}{V} (\frac{\partial V}{\partial T})|_P =-\frac{1}{\rho} (\frac{\partial \rho}{\partial T})|_P,
\end{equation}
\begin{equation}
\kappa_T = -\frac{1}{V} (\frac{\partial V}{\partial P})|_T =\frac{1}{\rho} (\frac{\partial \rho}{\partial P})|_T,
\end{equation}
are obtained as block averages from the set of runs. They are plotted in two panels of figure~\ref{fig7}.
One can conclude that the models used in simulations provide a qualitatively correct description
of these properties upon composition. However, in both  cases, the absolute values of
$\alpha_P$  and  $\kappa_T$ are overestimated in comparison with the experimental data.
The  TIP4P-2005-TraPPE model is better than the SPCE/E-TraPPE one. It is important
to mention that the $\kappa_T$ dependence on ethanol concentration 
from simulations captures the hydrophobic effect, in close similarity to the behavior of the density
shown previously in figure~\ref{fig1}b and in figure~\ref{fig3}a for the apparent molar volume of ethanol.

\begin{figure}[h]
\begin{center}
\includegraphics[width=6.5cm,clip]{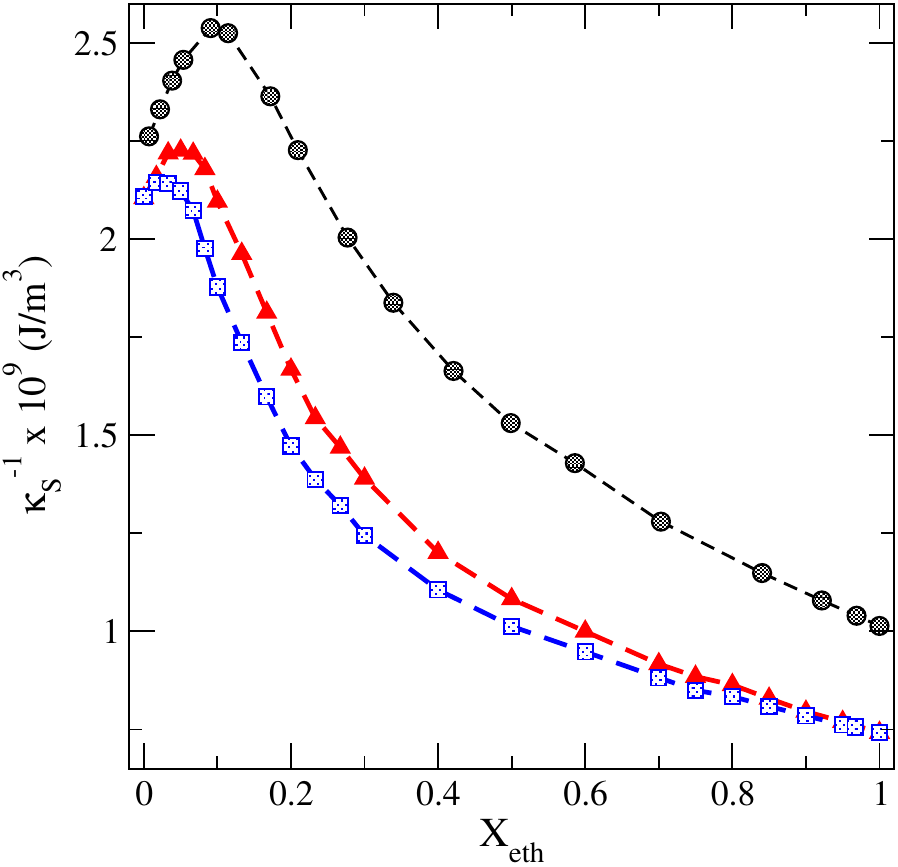}
\end{center}
\caption{(Colour online) A comparison of the composition dependence
of adiabatic bulk modulus, $\kappa_S^{-1}$,
with the experimental data. Circles are from~\cite{lara}.
\label{fig8}
}
\end{figure}

The adiabatic bulk modulus, which is the inverse of adiabatic compressibility, from 
simulations and experimental results is shown in figure~\ref{fig8}. These data are important because
the isoentropic compressibility is related to the speed of sound via Newton-Laplace equation, see, e.g.,
\cite{adiabatic}. Moreover, the isoentropic compressibility, $\kappa_S$ follows from the 
combination of other fluctuation-type properties, $\kappa_S = \kappa_T -T\alpha^2/\rho C_P$,
where $C_P$ is the constant pressure heat capacity. It was shown recently~\cite{adiabatic}
that simulations of TIP4P-2005-TraPPE model predict the composition dependence of the 
speed of sound for water-ethanol mixtures reasonably well (figure~\ref{fig1} of that reference). 
The maximum value of the speed of sound is reproduced at a slightly lower $X_{\text{eth}}$, 
in comparison with the experiment~\cite{lara}. Our comparison with the same set of experimental data
describes similar trends. In addition, we observe that the SPCE/E-TraPPE model captures
the maximum at even lower $X_{\text{eth}}$ than the TIP4P/2005-TraPPE one.

\begin{figure}[h]
\begin{center}
\includegraphics[width=6.5cm,clip]{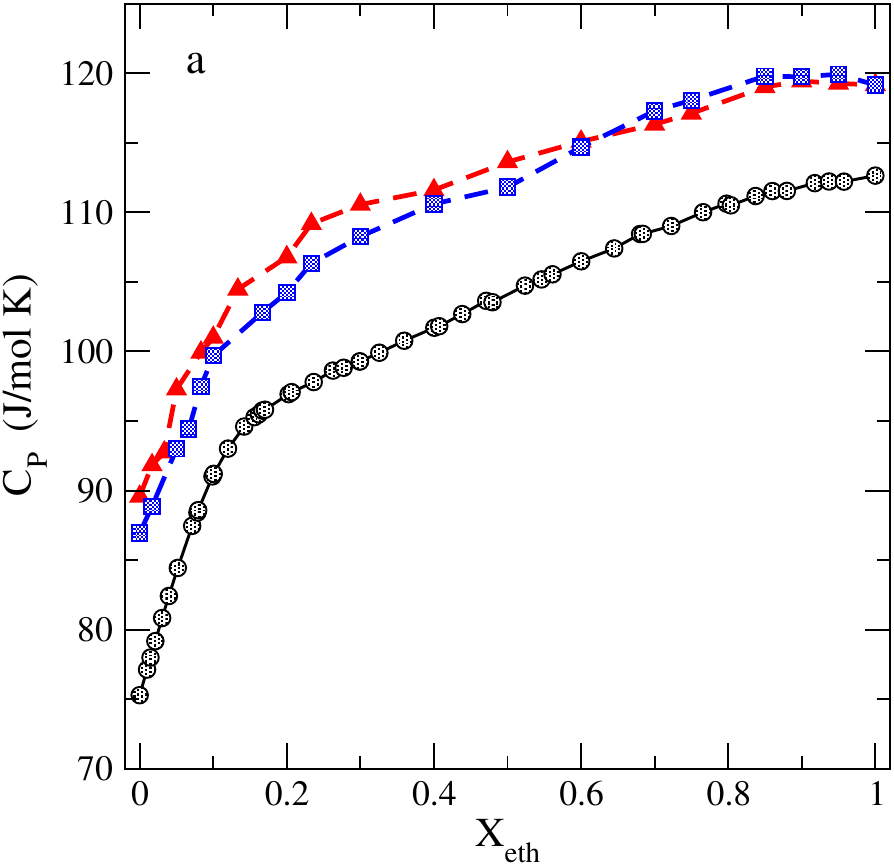}
\includegraphics[width=6.5cm,clip]{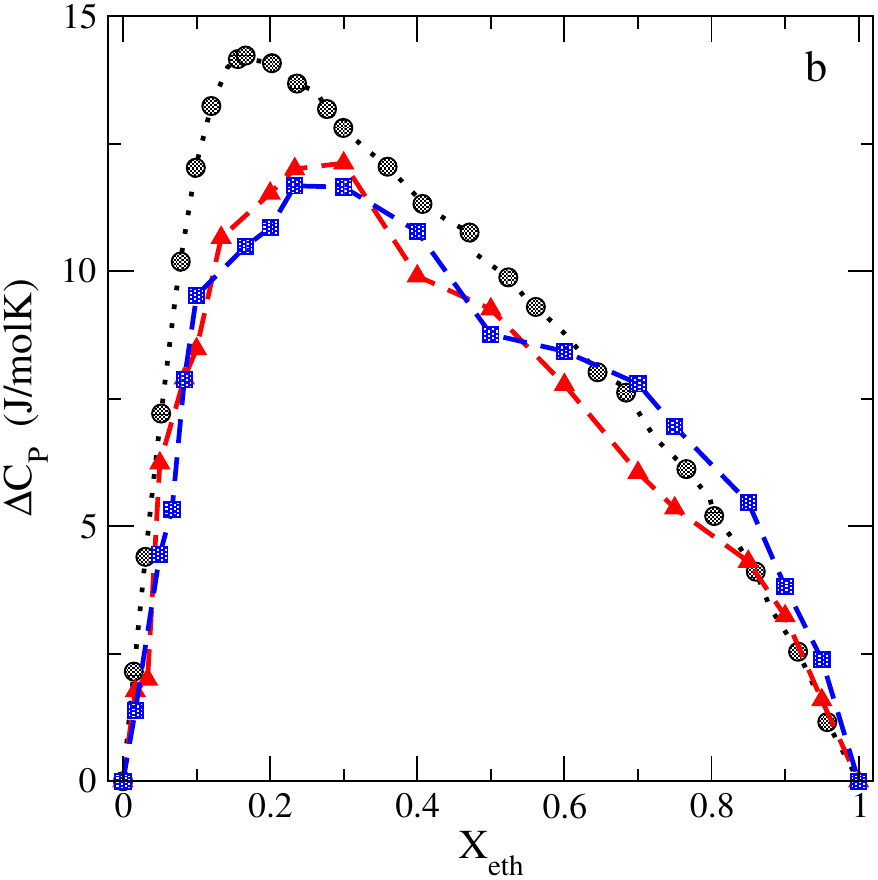}
\end{center}
\caption{(Colour online) Panels a and b: A comparison of the composition dependence 
of the heat capacity and the excess molar heat capacity
from simulations, with the experimental data \cite{benson} at 298.15~K.
The nomenclature of lines and symbols is the same as in figure~\ref{fig2}.
\label{fig9}
}
\end{figure}

We would like to conclude this subsection by the presentation of the simulation results for the 
constant pressure heat capacity. These are collected as averages for a set
of runs without applying any correction due to the density of states. The results
for the dependence of heat capacity and the excess heat capacity upon ethanol concentration
are given in figure~\ref{fig9}. The heat capacity values are overestimated in comparison with the experimental
result but the dependence of the excess heat capacity from simulations is reasonably good.
Both water models combined with TraPPE ethanol yield the results of comparable quality.

We would like to note here that some of these properties were explored by 
simulations of the TraPPE ethanol model combined with modified Fw-SPC water
model~\cite{cardona}. According to figure S1 of this work, the coefficient
of thermal expansion is closer to the experimental observations compared
to our data. On the other hand, again from figure~S1, it follows that
the minimum of isothermal compressibility at low ethanol concentrations 
is not captured appropriately. Finally, computer simulation predictions for the
heat capacity~\cite{cardona} are of similar quality as our results. 

\subsection{Self-diffusion coefficients of ethanol and water molecules}

\begin{figure}[h]
\begin{center}
\includegraphics[width=6.5cm,clip]{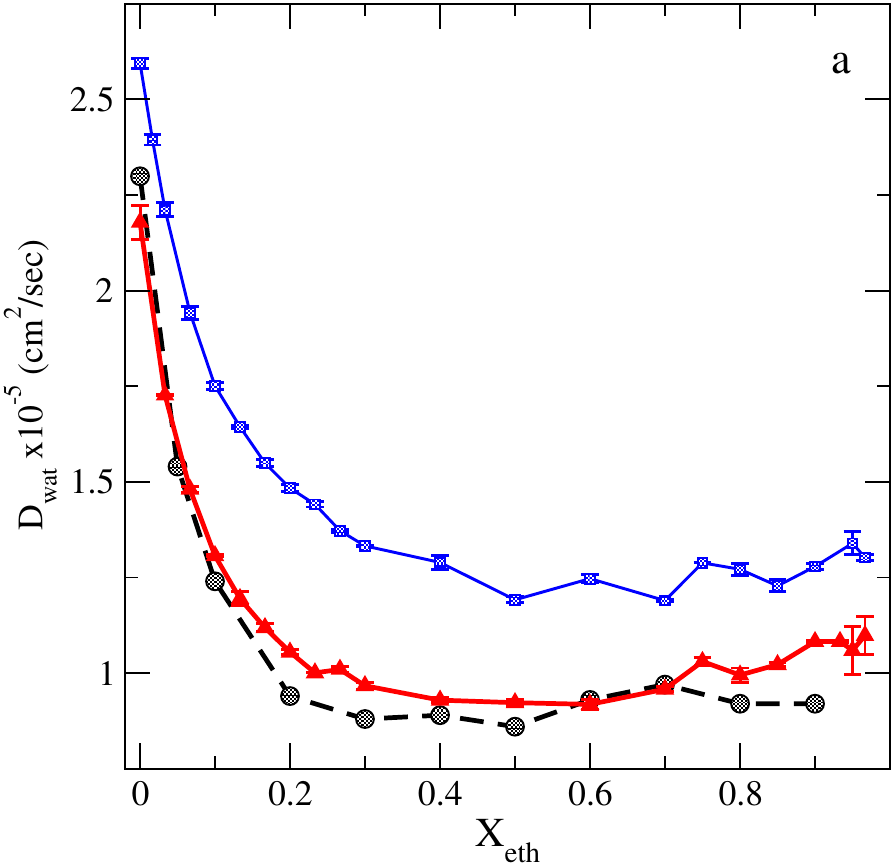}
\includegraphics[width=6.5cm,clip]{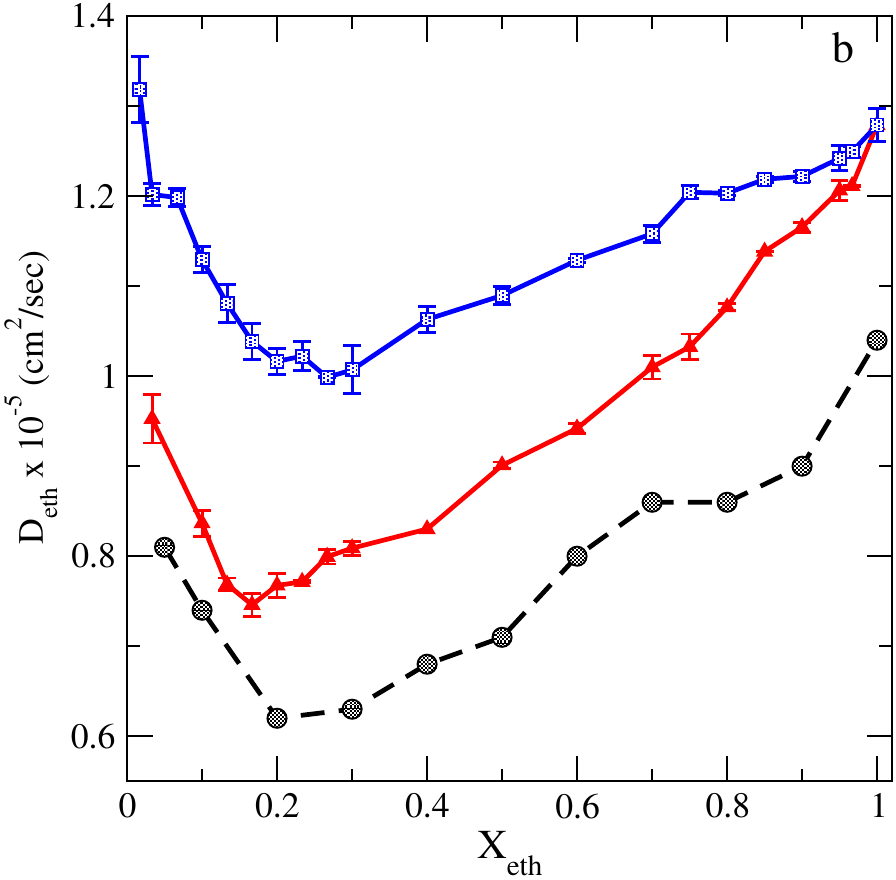}
\end{center}
\caption{(Colour online) Panels a and b: Composition dependence of
the self-diffusion coefficient  of water and ethanol, respectively,
in water-ethanol mixtures.
Experimental data are from \cite{price} (squares).
The nomenclature of lines and symbols as in figure~\ref{fig2}.
\label{fig10}
}
\end{figure}

One of the most popular properties to test models for binary mixtures is the 
self-diffusion coefficients of species. They can be obtained from the 
mean square displacement of particles or from calculations of the velocity
articulation functions.
We calculate the self-diffusion
coefficients, $D_i$ (i = w, eth), by the former route, via  the Einstein relation,
\begin{equation}
D_i =\frac{1}{6} \lim_{t \rightarrow \infty} \frac{\rd}{\rd t} \vert {\bf r}_i(\tau+t)-{\bf r}_i(\tau)\vert ^2,
\end{equation}
where  $\tau$ denotes the time origin. Default settings of GROMACS were used for the separation of
the time origins. The experimental data were taken from
\cite{price}. 
According to the experiments, the self-diffusion coefficient
of water species decreases in magnitude starting from pure water value (at $X_{\text{eth}}$ = 0) and reaches
minimum at $X_d \approx 0.5$, figure~\ref{fig10}a. Next, for higher values of  $X_{\text{eth}}$, $D_w$ does not change much. 
Apparently, the behavior of
$D_i(X_{\text{eth}})$ is determined by the evolution of density of the mixture and by hydrogen bonding
between all species, figure~\ref{fig10}b. 
Simulation predictions for $D_w$ show 
a satisfactory agreement with experimental trends in the entire composition interval,
if the TIP4P-2005-TraPPE model is used.
The SPC/E model overestimates the self-diffusion coefficient for water and consequently
the values for $D_w$ are overestimated for mixture at all compositions studied.
The shape of $D_w(X_{\text{eth}})$ curve is however similar to the experimental behavior and to the
TIP4P-2005-TraPPE model predictions.  The TraPPE model for ethanol does not 
describe the pure ethanol self-diffusion coefficient accurately. In consequence, only
the trends of behavior for $D_{\text{eth}}(X_{\text{eth}})$ are qualitatively correct. There is 
a minimum value at $X_{\text{eth}} \approx 0.2$ from the experimental data. The
TIP4P-2005-TraPPE shows a minimum value at $X_{\text{eth}} \approx 0.17$ whereas the
 SPC/E-TraPPE model --- at $X_{\text{eth}} \approx 0.27$. In general terms, the 
TIP4P-2005-TraPPE predictions are better.

It seems  necessary to confirm the present results by using alternative calculations 
via the velocity auto-correlation functions. In addition, it would be
profitable to  attempt calculations of the
relaxation times and possibly power spectra,  because
the experimental data are available in the literature. These issues would provide ampler 
insights into the appropriateness of the force fields to describe the dynamic properties.

\subsection{Static dielectric constant of ethanol-water mixtures}

\begin{figure}[h]
\begin{center}
\includegraphics[width=5.5cm,clip]{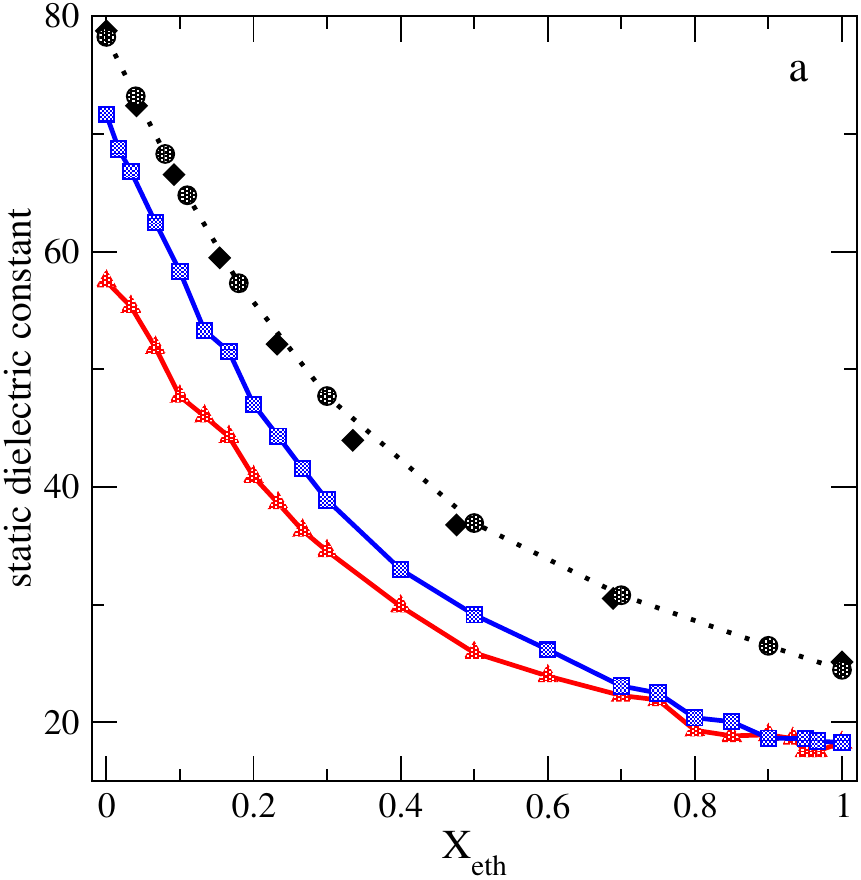}
\includegraphics[width=5.5cm,clip]{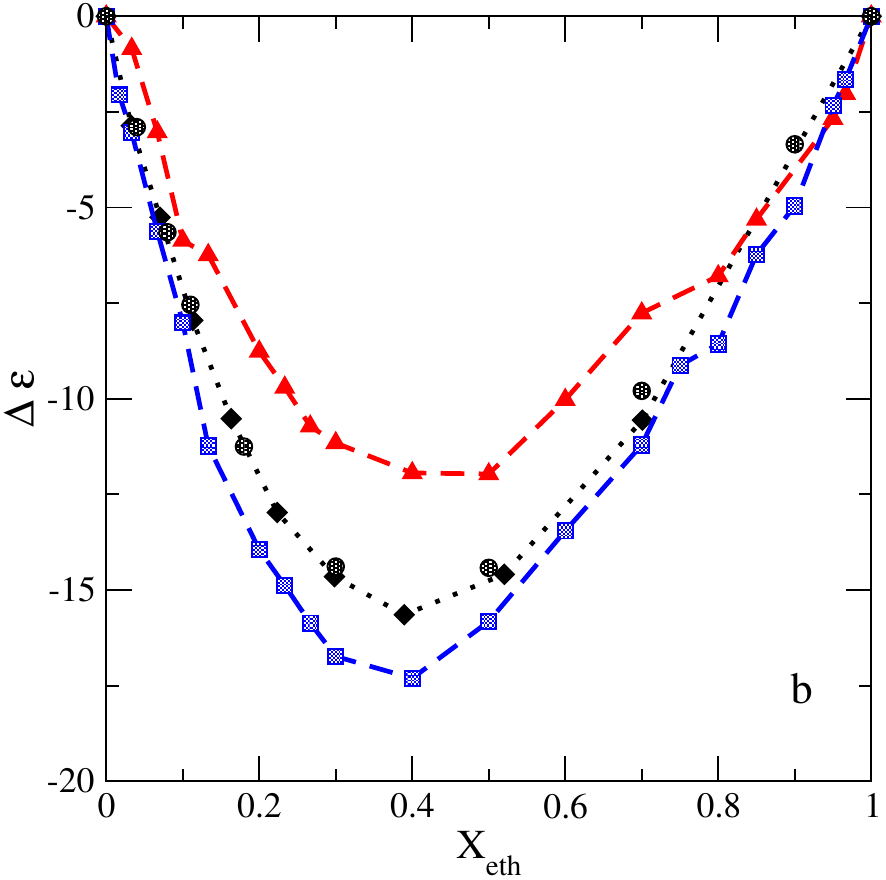}
\end{center}
\caption{(Colour online) Panels a and b: A comparison of the composition dependence
of the dielectric constant and the excess dielectric constant, respectively,
from simulations, with the experimental data (\cite{wolf} --- diamonds) at 298.15~K.
The nomenclature of lines and symbols is the same as in figure~\ref{fig2}.
Another set of data is from \cite{buchner} (circles).
\label{fig11}
}
\end{figure}

The study of dielectric properties of water-alcohol mixtures represents a
wide area of research, see, e.g., quite recent contribution~\cite{cardona}, 
as an example concerned with some interesting issues for water-ethanol mixtures.
Here, we would like to focus solely on the calculation of the static dielectric constant.
The long-range, asymptotic behavior of correlations between molecules
possessing a dipole moment is described by the dielectric constant, $\varepsilon$.
It is calculated from the time-average of the fluctuations of the total
dipole moment of the system~\cite{martin} as follows,

\begin{equation}
\varepsilon=1+\frac{4\piup}{3k_\text{B}TV}(\langle\bf M^2\rangle-\langle\bf M\rangle^2),
\end{equation}
where $k_\text{B}$ is the Boltzmann constant and $V$ is the simulation cell volume.
Technical aspects of calculations were commented by us in several 
publications, see, e.g., \cite{aguilar}.
The experimental data are taken from~\cite{wolf,buchner}.
The dielectric constant, $\varepsilon$,  and the excess dielectric constant,
$\Delta \varepsilon$ (defined similarly to excess molar volume or enthalpy, see section~\ref{sec3.2})
are plotted in figure~\ref{fig11}. The dielectric constant decreases from a high value for water to a
much lower value for pure ethanol, figure~\ref{fig11}a. 

Computer simulations of the models in question underestimate $\varepsilon$ in the
entire composition interval. On the water-rich side, the problem is that both
water models in question underestimate $\varepsilon$ of pure water. 
This issue can be solved by considering the TIP4P/$\varepsilon$ model~\cite{alejandre}.
On the ethanol-rich side, the problem is that the TraPPE ethanol model
underestimates $\varepsilon$ of pure ethanol. The curve of similar quality
follows from the simulations of TraPPE ethanol combined with Fw-SPC water,
which is given in figure~6a of \cite{cardona}. A better ethanol model
is required to mitigate this deficiency. One possibility is to involve
the primary alcohols model designed in the laboratory of J. J. de Pablo~\cite{nerd}.
We are aware of the successful parametrization of the dielectric constant
of propanol-water mixtures, \cite{mendez}, by using this model.
For ethanol this issue has not been solved so far.

In spite of inaccuracy of the absolute values for the dielectric constant, the
shape of $\varepsilon(X_{\text{eth}})$ is quite reasonable, figure~\ref{fig11}b. 
The maximum deviation from ideality is observed at $X_{\text{eth}} \approx 0.4$, 
as in experiments. The  SPC/E-TraPPE results are closer to the experimental 
curve than the TIP4P/2005-TraPPE ones. Still, in general terms the agreement
of simulation data and experiment can be termed as satisfactory.

\subsection{Surface tension of ethanol-water mixtures on composition}

Our final remarks concern the behavior of the surface tension of ethanol-water
mixtures. This research topic has long history as documented in~\cite{tarek}.
Recent studies refer to experimental
characterization of this liquid-vapor interface~\cite{hyde,kirschner} 
and its study using molecular dynamics simulations~\cite{camp}.

\begin{figure}[h]
\begin{center}
\includegraphics[width=6.0cm,clip]{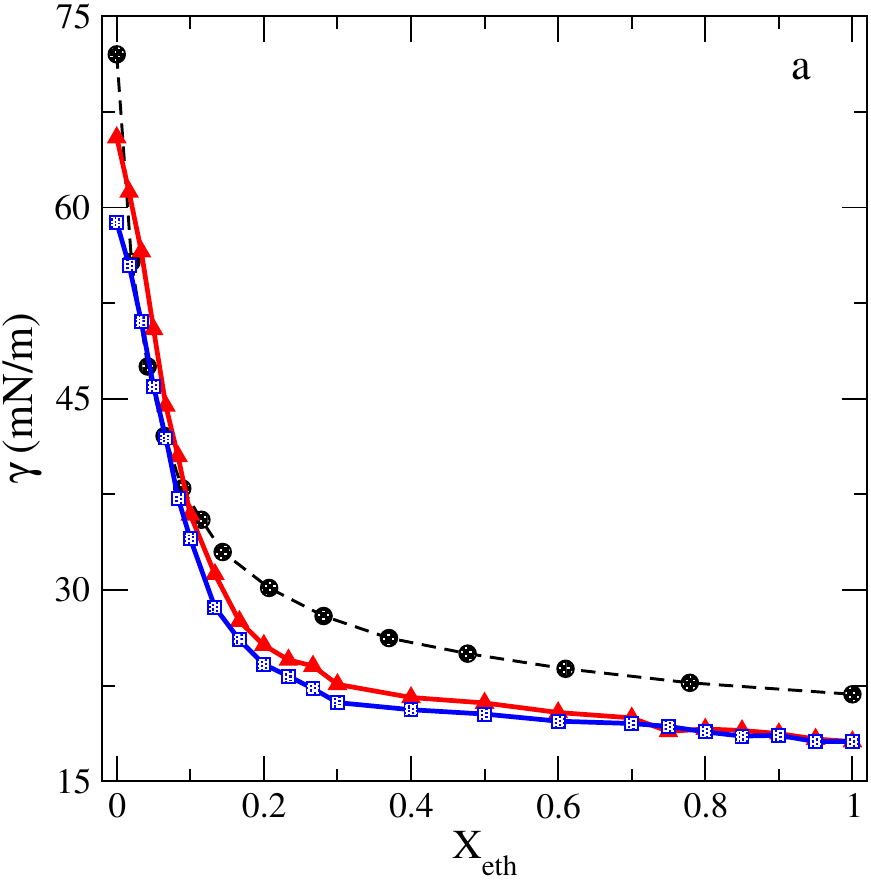}
\includegraphics[width=6.0cm,clip]{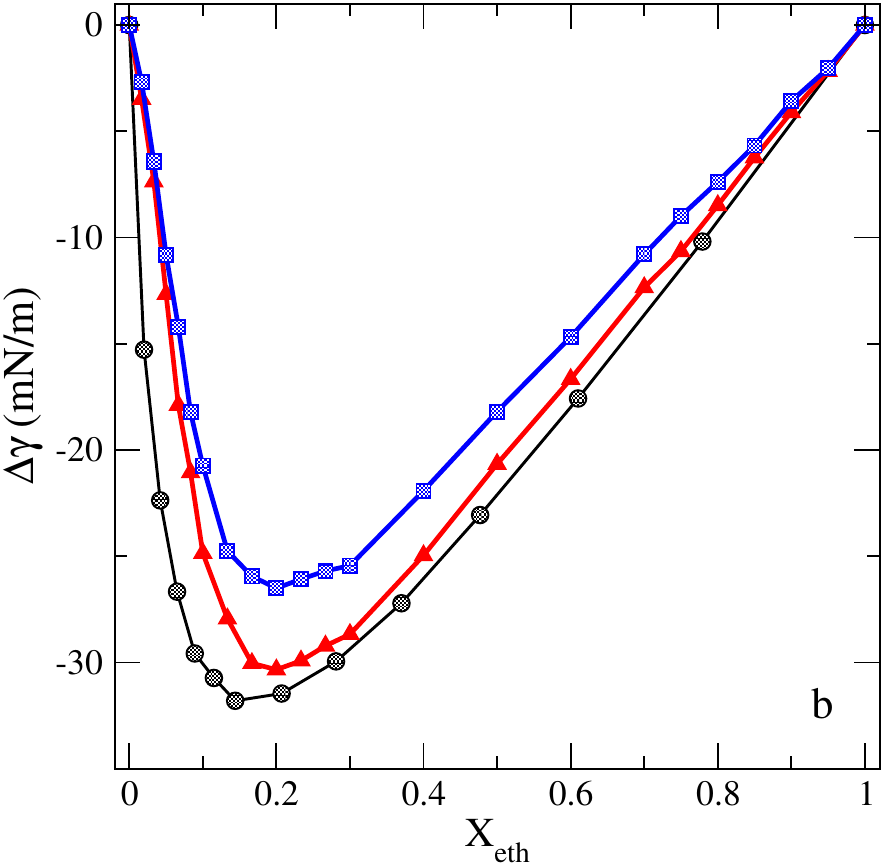}
\end{center}
\caption{(Colour online) Panels a and b: A comparison of the composition dependence
of the surface tension and the excess surface tension, respectively,
from simulations, with the experimental data \cite{vazquez} at 298.15~K.
The nomenclature of lines and symbols is the same as in figure~\ref{fig2}.
\label{fig12}
}
\end{figure}

The  surface tension calculations at each
composition were performed by taking the final configuration of particles
from the isobaric run. However, it is well known that the cutoff distance should be
increased to obtain accurate values for surface tension. After short trials,
we   chose, $r_c = 1.4$ nm, in what follows.
Next, the box edge along $z$-axis was extended by a factor of 3,
generating a rectangular box with liquid slab
and two liquid-mixture-vacuum interfaces in the $x-y$ plane,
in close similarity to the procedure used in~\cite{vanderspoel}.
The total number of molecules is sufficient to yield an area of the
$x-y$ face of the liquid slab sufficiently big. The elongation of the liquid slab along
$z$-axis is satisfactory as well. The executable file was modified by deleting
a fixed pressure condition preserving the $V$-rescale thermostatting with the same parameters
as in the NPT runs. Other corrections were not employed.

The values for the surface tension, $\gamma$, follow from the combination of the time
averages for the components of the pressure tensor,
\begin{equation}
\gamma = \tfrac{1}{2} L_z \langle[P_{zz}-\tfrac{1}{2}(P_{xx}+P_{yy})]\rangle,
\end{equation}
where $P_{ij}$ are the components of the pressure tensor along $i,j$ axes, and  $\langle\ldots\rangle$
denotes the time average.
We performed a set of runs at a constant volume, each piece of 10 ns, and
obtained the result for $\gamma$ by taking the block average.
The experimental results were taken from \cite{vazquez}.
The data show that the surface tension rapidly decreases from the pure water
value at $X_{\text{eth}}=0$  till $X_{\text{eth}} \approx0.25$. Next, at higher $X_{\text{eth}}$, the values
for $\gamma$ decrease more slowly, the curve behaves almost linearly in that interval
of compositions, panel a of figure~\ref{fig12}. 
At $X_{\text{eth}}=0$, we have $\gamma \approx 65.5$ for TIP4P/2005 water and
$\gamma \approx 58.8$ for SPC/E water model without long-range correction,
in agreement with data 65.3 and 60.2, reported in \cite{surftip}, respectively. 
For TraPPE ethanol, $X_{\text{eth}}=1$, we obtained $\gamma \approx 18.1$, in close similarity
to the points in figure~7 of \cite{obeidat}. 
In the interval of not small ethanol concentration, $X_{\text{eth}} > 0.15$,
the absolute values for $\gamma$ from simulations are underestimated compared to experimental data. 
It is difficult to expect that inclusion of long-range corrections would
mitigate this problem. Rather, one should search for parametrization of the ethanol model.

\begin{figure}[h]
\begin{center}
\includegraphics[width=5cm,clip]{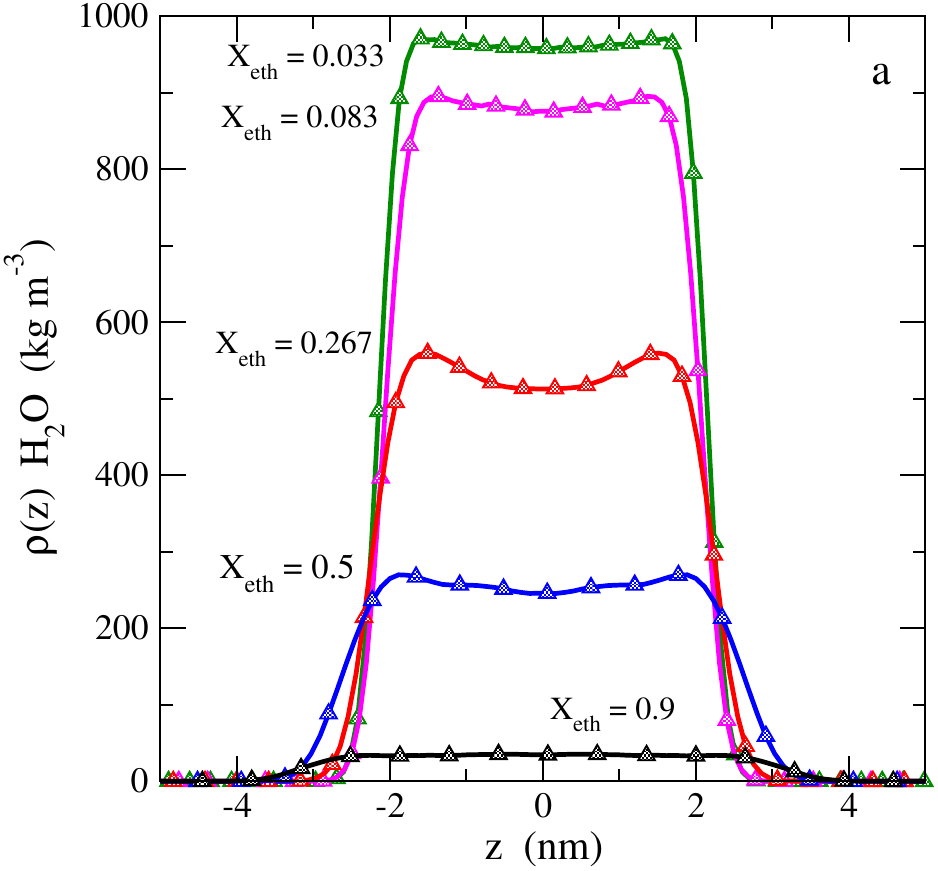}
\includegraphics[width=5cm,clip]{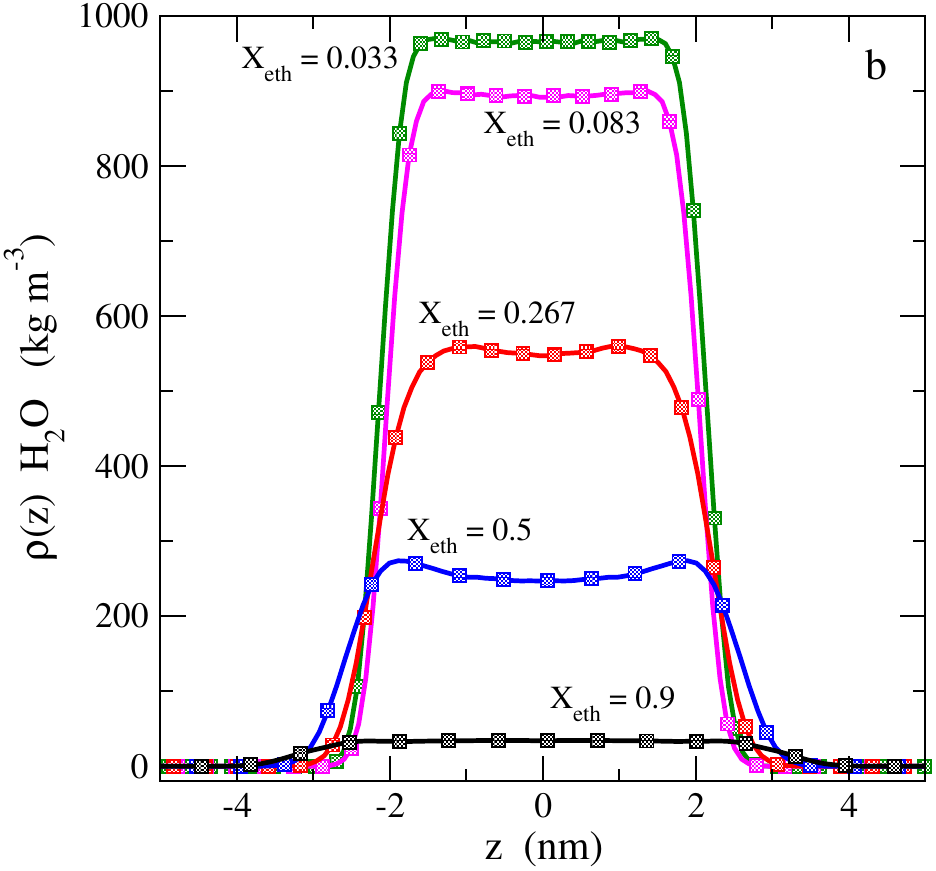} \\ 
\includegraphics[width=5cm,clip]{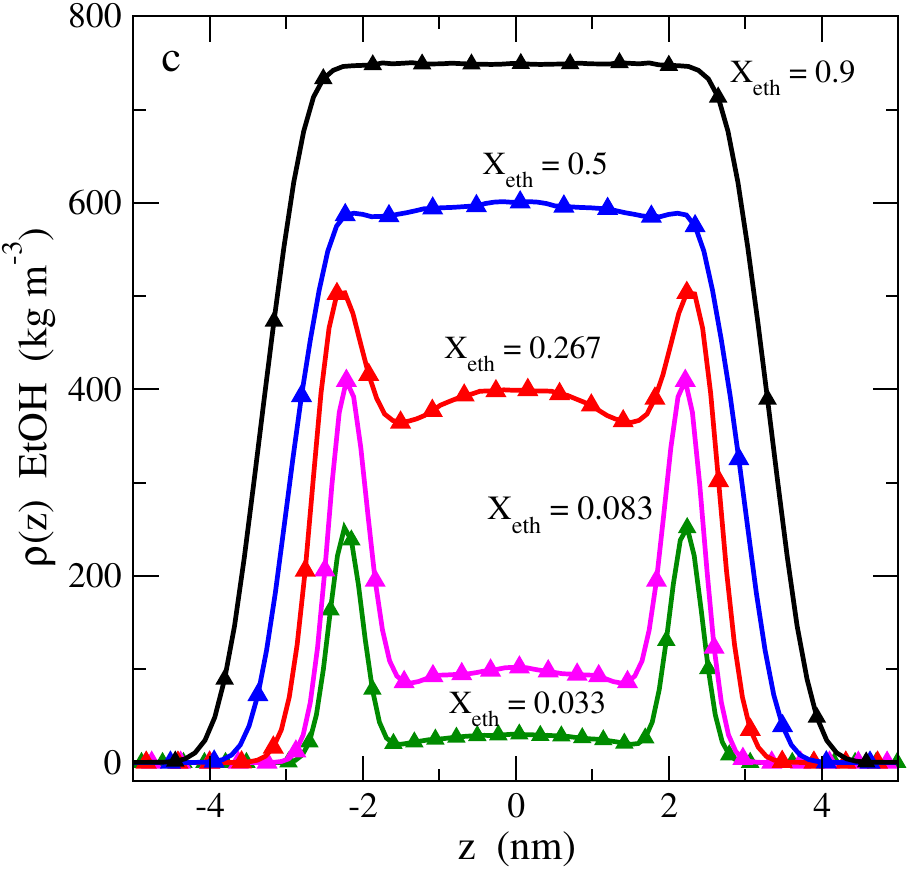}
\includegraphics[width=5cm,clip]{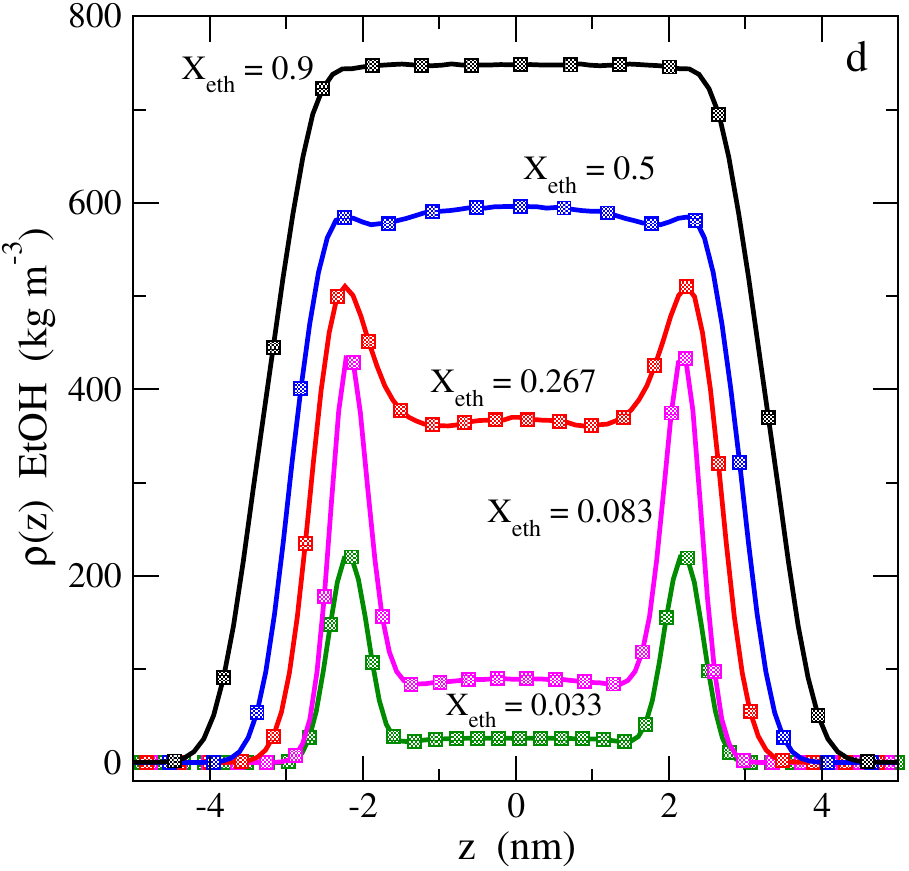}
\end{center}
\caption{(Colour online) Mass density profiles of water species through vapor-liquid interface
on composition for TIP4P-2005-TraPPE (panel a) and SPC/E-TraPPE
model (panel b). Density profiles of ethanol species
for TIP4P-2005-TraPPE (panel c) and SPC/E-TraPPE model (panel d).
Triangles refer to TIP4P-2005-TraPPE profiles whereas
the squares are for SPC/E-TraPPE model.
\label{fig13}}
\end{figure}

The excess surface tension from experiment
exhibits minimum at $X_{\text{eth}} \approx$0.17. Two models in question behave similarly
for water-rich mixtures. The decay of $\gamma$ is a bit better reproduced by 
the TIP4P-2005-TraPPE model, in comparison with SPC/E -TraPPE one.
It is of interest to explore the structure of the interface in terms of density profiles 
of the species. Previously, these profiles were 
reported in~\cite{tarek,camp}. Namely, in \cite{camp}, the dependence of the
density profiles on water-ethanol mixture composition was studied.
We use a similar type of presentation of the profiles in figure~\ref{fig13}.
Besides, the properties of the air/solution interface were examined by Tarek et al.~\cite{tarek} 
at single, 0.1~M,   composition of ethanol-water mixture.

\begin{figure}[h]
\begin{center}
\includegraphics[width=6.5cm,clip]{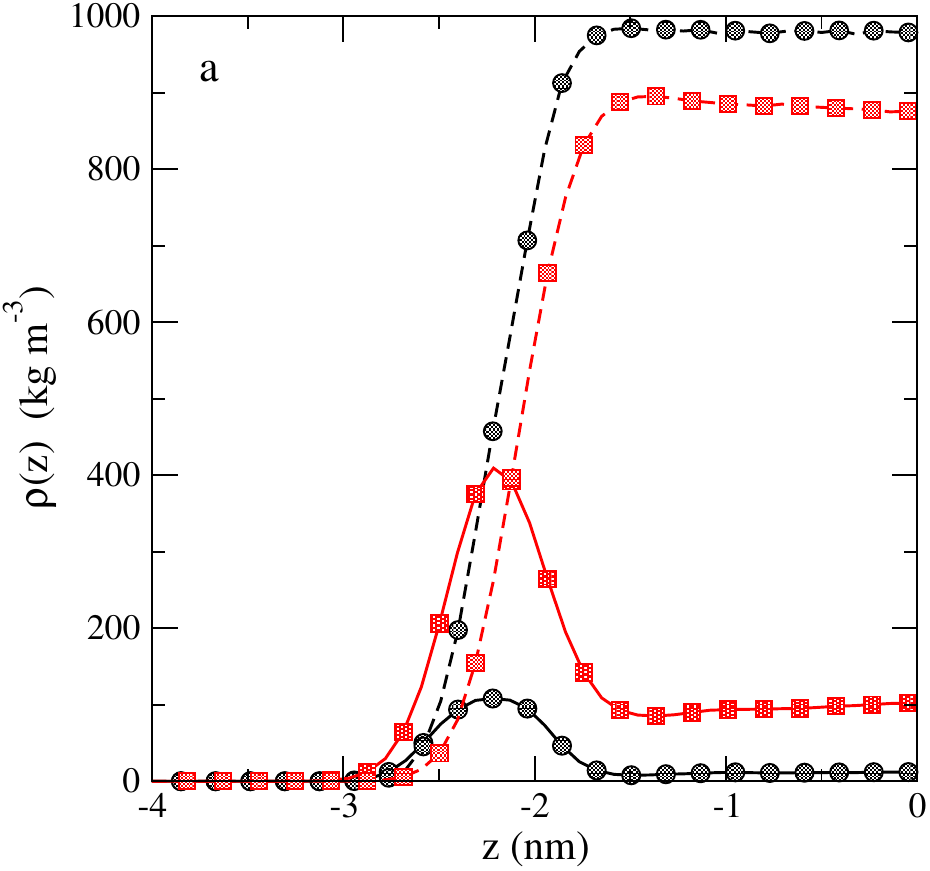}
\includegraphics[width=6.5cm,clip]{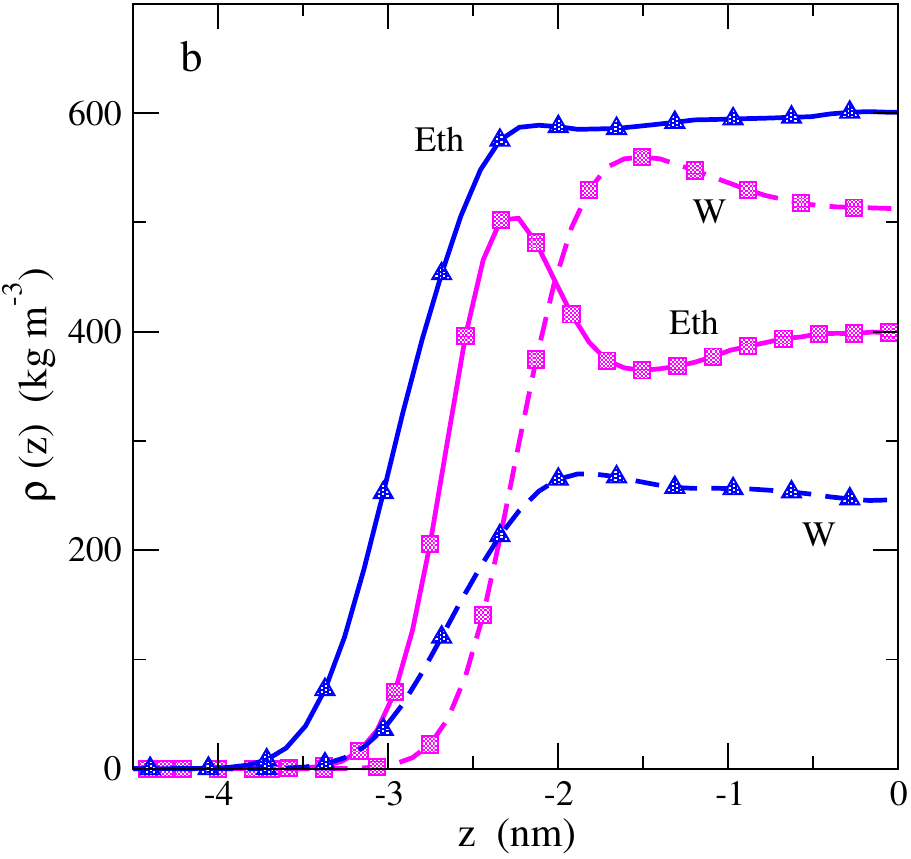}
\end{center}
\caption{(Colour online) A comparison of the mass density profiles of the species 
through liquid-vapor interface with
TIP4P/2005-TraPPE model  at different compositions. In panel a,
$X_{\text{eth}}=0.0167$ (black lines and circles),
$X_{\text{eth}}=0.0833$ (red lines and squares). In panel b,
$X_{\text{eth}}=0.267$ (magenta lines and squares) and
$X_{\text{eth}}=0.5$ (blue lines and triangles). Profiles of water and of ethanol species 
are given by dashed and solid lines, respectively.
\label{fig14}
}
\end{figure}

The left-hand panels of figure\ref{fig13} refer to TIP4P/2005-TraPPE model, whereas
two right-hand panels show similar results but for SPC/E-TraPPE model.
At a low concentration of ethanol species in the bulk phase, $X_{\text{eth}} = 0.03$,
one can see that the ethanol molecules segregate from the bulk to the 
interface (panel c of figure~\ref{fig13}) and form a monolayer. In order
to explain the changes of the location of the monolayer, we refer to the
panel a of figure\ref{fig14}. At a very low $X_{\text{eth}}$,  $X_{\text{eth}} = 0.0167$, the interface composition is
dominated by water species. However, already at $X_{\text{eth}} = 0.0833$, segregation
of ethanol molecules from the bulk becomes stronger, so that 
these species overcome water concentration within the interface region.

A certain degree of  depletion of water molecules from the bulk to the interfaces is observed
in the concentration interval, $0.0833 <  X_{\text{eth}} < 0.5$,  according to
the TIP4P/2005-TraPPE model (panel a of figure~13). By contrast,
these trends are much weaker within the SPC/E-TraPPE model (panel b of figure~13).  
At a higher ethanol concentration in the bulk, e.g., $X_{\text{eth}} = 0.5$,
the interface composition is dominated by the ethanol molecules, figure~\ref{fig13}a and figure~\ref{fig14}b. 

This set of results complement the previous observations made in \cite{camp,tarek}
concerning composition evolution of the liquid-vapor interface of water-ethanol mixtures.

\section{Summary and conclusions}

Some time ago, we  undertook rather comprehensive studies of the 
composition evolution of the properties of 
water-methanol mixtures~\cite{galicia,mario,galicia2}.
In the present work, we explore water-ethanol  systems
with the same methodology but using the non-polarizable, united atom ethanol TraPPE model
in conjunction with the TIP4P-2005 and SPC/E water models. 

A set of novel findings concerning a wide range of properties of these models
were obtained and discussed in comparison to experimental data. Namely,
we obtained the excess apparent molar
volumes and excess partial molar enthalpies. 
Analyses of these properties  permitted us to capture an anomalous behavior
observed at a low ethanol molar fraction in agreement with experimental results.
In addition, we described the composition dependence of the properties coming from 
calculations of fluctuations.
The self-diffusion coefficients of species were obtained.
Next, the static dielectric constant was calculated and
compared with experimental results. Our final focus was on elucidation of the behavior of
the surface tension on the composition of water-ethanol mixtures and of the excess
surface tension. The density profiles of the species through the liquid-vapor interface
are discussed. All the trends were compared with experimental data.
A detailed validation of the predictions of
the properties resulting from a set of models for water-ethanol  solutions
permits to make conclusions w.r.t. their applicability.
We have not observed the trends that contradict the experimental findings.
In general terms, the TIP4P/2005-TraPPE model seems to be superior in 
comparison with SPC/E-TraPPE model. This latter model is better for the
description of the dielectric constant only. However, the TIP4P/2005
can be replaced by TIP4P$\varepsilon$ model to mitigate the problem.
Various excess mixing properties are better decribed than the corresponding 
absolute values. Still, it seems that the improvement of agreement with
experimental data can be reached only by designing  a more accurate ethanol model.

Finally, we would like to enlist a few missing elements to extend our
knowledge of the properties of water-ethanol mixtures.
It would be profitable to explore various velocity
auto-correlation functions to enhance the understanding of dynamic properties, as well as
the frequency dependent dielectric constant
to  better understand the dielectric properties. We were unable to extract useful pieces of
information from the Kirwood factors, for the moment.
As it was mentioned in the introduction, all the issues concerned with the
composition evolution of the microscopic structure and hydrogen bonds network
will be reported elsewhere.

Moreover, a detailed comparison of the predictions coming out from united atom and
all atom force fields for water-alcohol mixtures is still missing at present.
Recent, quite general analyses of the mixing trends were given in \cite{idrissi}
by using Monte Carlo simulations.
Qualitative agreement of the results of models with a different degree of sophistication
with experimental behavior was observed.
Nevertheless, these findings require an extension for specific systems in future work.  
On the other hand, some recent efforts were focused on the improvement
of the description of thermodynamic and other properties by multi-step parametrization 
of united atom models for simple alcohols with non-polarizable water models~\cite{edgar}.
Again, extensive additional work is needed to reach definite conclusions
concerning the accuracy of the constructed force fields.

It is worth mentioning that
the solvation of complex molecules in water-ethanol mixtures was the subject of
several simulations and experimental studies, see, e.g.,
\cite{camp,bagchi1,bagchi2,brasil}.  
Some important aspects within this research area are under study in our laboratory.
Namely, along the lines considered in the report from Scotch whisky research institute~\cite{camp}, 
we intend to analyze the trends of the smell of beverages of Mexican origin, see, e.g.,~\cite{nuria},
by using computer simulation techniques.  Concerning the experimental
work focused on the properties of molecules that are promising for medical applications~\cite{brasil}, 
our present interest is in clustering of curcumin
molecules in water-ethanol mixtures, as an extension of  our recent 
contribution~\cite{patsahan}. Progress along these research lines will be reported 
elsewhere.

\section*{Acknowledgements} O.P. acknowledges helpful discussions with Dr. Laszlo Pusztai
concerning experimental measurements of the structure of water-ethanol mixtures.

\newpage
\ukrainianpart

	\title{Моделювання суміші води з етанолом методом молекулярної динаміки. I. Композиційні тенденції в термодинамічних властивостях}

\author{Д. Бенавiдес Батiста\refaddr{label1},
	М. Агілар\refaddr{label2}, 
	О. Пізіо\refaddr{label2}}
\addresses{
	\addr{label1} Інститут фундаментальних наук та техніки, Автономний університет штату Ідальго, Пачука-де-Сото, Ідальго 42039, Мексика
	\addr{label2}Інститут хімії, Національний автономний університет Мексики, 
	Сіркуіто Екстеріор, 04510, Мексика}
\makeukrtitle
\begin{abstract}
	
	Досліджено композиційну залежність доволі широкого діапазону властивостей рідких сумішей води з етанолом за допомогою комп'ютерного моделювання методом ізобарно-ізотермічної молекулярної динаміки. Розглянуто неполярну модель для молекули етанолу з бази даних TraPPE у поєднанні з моделями води TIP4P-2005 і SPC/E. У наших розрахунках ми обмежуємося значеннями атмосферного тиску 0.1013 МПа та кімнатної температури 298.15~K. Описано композиційні тенденції у поведінці густини, надлишкового об'єму змішування та видимих молярних об'ємів. Також досліджено надлишкову ентальпію змішування і парціальні молярні ентальпії компонент суміші. Крім того, ми досліджуємо коефіцієнт ізобарного теплового розширення, ізотермічну теплоємність, адіабатичний модуль об'ємного стиску та теплоємність при постійному тиску. Також розраховані коефіцієнти самодифузії частинок, статична діелектрична проникність і поверхневий натяг. Ми сподіваємось отримати хороше уявлення про особливості змішування компонент суміші при зміні молярної частки етанолу. Якість прогнозних оцінок критично оцінюється шляхом детального порівняння з експериментальними даними. Насамкінець, обговорюються необхідні вдосконалення схеми моделювання.
	\keywords молекулярна динаміка, водні суміші етанолу, поверхневий натяг, діелектрична проникність, парціальні молярні об'єми
\end{abstract}
\end{document}